\pdfoutput=1
\documentclass[11pt,letterpaper]{article}

\usepackage[letterpaper,margin=1in]{geometry}
\usepackage[T1]{fontenc}
\usepackage{lmodern}
\usepackage{microtype}
\usepackage{setspace}
\usepackage{parskip}
\usepackage{amsmath,amssymb}
\usepackage{booktabs,makecell,array,longtable}
\newcolumntype{L}[1]{>{\raggedright\arraybackslash}p{#1}}
\usepackage{graphicx}
\usepackage{xcolor}
\usepackage{enumitem}
\usepackage{caption}
\usepackage{titlesec}
\usepackage[section]{placeins}
\usepackage{fancyvrb}
\usepackage{xspace}
\usepackage{tikz}
\usetikzlibrary{arrows.meta,positioning,fit,calc,backgrounds}
\usepackage[skins,breakable]{tcolorbox}
\usepackage{xurl}
\usepackage[colorlinks=true,linkcolor=blue!45!black,citecolor=blue!45!black,urlcolor=blue!45!black]{hyperref}
\usepackage[capitalise,noabbrev]{cleveref}
\input{glyphtounicode}


\titleformat{\section}{\sffamily\bfseries\Large}{\thesection}{0.7em}{}
\titlespacing*{\section}{0pt}{2.6ex plus 0.6ex minus 0.2ex}{1.2ex plus 0.2ex}
\titleformat{\subsection}{\sffamily\bfseries\large}{\thesubsection}{0.7em}{}
\titlespacing*{\subsection}{0pt}{2.2ex plus 0.5ex minus 0.2ex}{0.9ex plus 0.1ex}
\titleformat{\subsubsection}{\sffamily\bfseries\normalsize}{\thesubsubsection}{0.7em}{}
\titlespacing*{\subsubsection}{0pt}{1.8ex plus 0.4ex minus 0.2ex}{0.6ex plus 0.1ex}

\setlist{itemsep=0.35em,topsep=0.3em,parsep=0.2em}
\crefname{section}{Section}{Sections}
\crefname{subsection}{Section}{Sections}
\crefname{subsubsection}{Section}{Sections}
\crefname{appendix}{Appendix}{Appendices}
\crefname{equation}{Equation}{Equations}
\creflabelformat{equation}{#2(#1)#3}
\crefname{table}{Table}{Tables}
\crefname{figure}{Figure}{Figures}

\newtcolorbox{shows}{enhanced,unbreakable,colback=black!4,colframe=black!55,boxrule=0pt,leftrule=2.2pt,
  arc=0pt,left=9pt,right=9pt,top=5pt,bottom=5pt,before skip=9pt,after skip=11pt,
  before upper={\sffamily\bfseries\small What this shows\par\vspace{2pt}\rmfamily\mdseries\normalsize}}
\newtcolorbox{answer}[1]{enhanced,unbreakable,colback=black!9,colframe=black!70,boxrule=0pt,leftrule=2.2pt,
  arc=0pt,left=9pt,right=9pt,top=5pt,bottom=5pt,before skip=9pt,after skip=13pt,
  before upper={\sffamily\bfseries\small #1\par\vspace{2pt}\rmfamily\mdseries\normalsize}}
\newtcolorbox{statement}[1]{enhanced,unbreakable,colback=white,colframe=black!60,boxrule=0.5pt,
  arc=1pt,left=9pt,right=9pt,top=5pt,bottom=5pt,before skip=9pt,after skip=11pt,
  before upper={\sffamily\bfseries\small #1\par\vspace{2pt}\rmfamily\mdseries\normalsize}}

\newcommand{\tablesetup}{\small\setlength{\tabcolsep}{4.5pt}}

\newcommand{\NgenCases}{1,360}
\newcommand{\NfeRuns}{30}
\newcommand{\NengMutants}{350}

\newcommand{\NneHookPctExt}{77}

\newcommand{\NpkgCases}{186}
\newcommand{\NpkgMalformed}{13}
\newcommand{\NpkgDrivable}{180}
\newcommand{\NpkgRuns}{5}
\newcommand{\NextRuns}{3}
\newcommand{\NlatCases}{77}
\newcommand{\NlatRuns}{5}
\newcommand{\NevidPairsExt}{15,146}
\newcommand{\NhookBigintCases}{5}
\newcommand{\NextWallMin}{12}
\newcommand{\NengHours}{4.2}

\newcommand{\NengineOnly}{6}
\newcommand{\NfeApplicable}{2,660}
\newcommand{\NfeCaught}{2,660}
\newcommand{\NdropArgs}{217}
\newcommand{\NdropArgsX}{120}
\newcommand{\NdropArgsEvOnly}{97}
\newcommand{\NextCases}{1,546}
\newcommand{\NlatGroups}{3,850}
\newcommand{\NlatDisagree}{0}
\newcommand{\NlatCallsPerRun}{3,080}
\newcommand{\NlatPSdk}{39.5}
\newcommand{\NlatPMcp}{91.8}
\newcommand{\NlatPEgress}{54.7}
\newcommand{\NlatPHook}{43.9}
\newcommand{\NlatDSdk}{39.8}
\newcommand{\NlatDMcp}{89.3}
\newcommand{\NlatDEgress}{52.7}

\newcommand{\NeqExtPerRun}{1,865}
\newcommand{\NwireCompared}{1,540}

\newcommand{\NneMcpAdapter}{878}
\newcommand{\NneMcpInherent}{317}
\newcommand{\NneMcpAll}{1,195}

\newcommand{\NneEgressInherent}{367}
\newcommand{\NneEgressAll}{367}
\newcommand{\NneHookAdapter}{857}
\newcommand{\NneHookInherent}{331}
\newcommand{\NneHookAll}{1,188}
\newcommand{\NpkgFailRules}{25}
\newcommand{\NextDrivable}{1,540}
\newcommand{\NfeRunsAll}{31}
\newcommand{\NmSplitMedianCases}{127}
\newcommand{\NmSplitOneCase}{17}
\newcommand{\NlatSpreadMax}{2.1}
\newcommand{\NfsDec}{13}
\newcommand{\NfsEv}{8}
\newcommand{\NfsErr}{55}

\newcommand{\NfsSurv}{73}
\newcommand{\NfeMutants}{149}

\newcommand{\NmOpaSurv}{45}

\newcommand{\NmSharedSplit}{51}

\newcommand{\NmSharedExpOnly}{5}
\newcommand{\NmSharedSurv}{38}

\newcommand{\NmSharedTotal}{108}
\newcommand{\NmAllSplit}{228}
\newcommand{\NmAllErr}{9}
\newcommand{\NmAllMutOnly}{7}
\newcommand{\NmAllExpOnly}{5}
\newcommand{\NmAllSurv}{86}

\newcommand{\NmAllComp}{8}

\newcommand{\NmOpaSurvRoleKnown}{21}
\newcommand{\NmLive}{343}
\newcommand{\NmSplitPct}{66}
\newcommand{\NmSurvPct}{25}
\newcommand{\NmFlagPkg}{243}
\newcommand{\NmFlagExt}{249}
\newcommand{\NmFlagGenOnly}{6}
\newcommand{\NlatWarmup}{40}
\newcommand{\NlatRounds}{10}
\newcommand{\NdtCases}{131}
\newcommand{\NdtCross}{46}

\newcommand{\NrefOeMissed}{2}

\newcommand{\NliveMutants}{4}
\newcommand{\NfaultHours}{0.9}
\newcommand{\NfeSrcHours}{3.7}
\newcommand{\NengEvaluated}{1,539}
\newcommand{\sys}{Oathpost}
\newcommand{\code}[1]{\texttt{#1}}

\makeatletter
\newcommand{\zenododoi}[1]{%
  \def\pxzen@doi{#1}\def\pxzen@none{TO-BE-ADDED}%
  \ifx\pxzen@doi\pxzen@none\else
    \ The same archive is also deposited at Zenodo, DOI~\href{https://doi.org/#1}{\nolinkurl{#1}}.%
  \fi}
\makeatother

\hypersetup{pdftitle={One Policy, Four Interception Points: Differential Testing of Agent Governance},
  pdfauthor={Animesh Shaw}}

\title{\sffamily\bfseries One Policy, Four Interception Points:\\[0.3ex]
Differential Testing of Agent Governance}
\author{Animesh Shaw\\[0.3ex] \small Golden Gate University, San Francisco, CA, USA\\
\small\texttt{ashaw406@my.ggu.edu}}
\date{}

\begin{document}
\setstretch{1.15}
\maketitle

\begin{abstract}
\noindent
Governance layers for AI agents decide, before an agent's action runs, whether it may run, and
record why. They can sit at four interception points: inside the agent's process (through a software development kit, SDK), in front of
a Model Context Protocol (MCP) server, on the network path (egress proxy), or in a coding agent's
pre-tool hook. Operators expect one policy to mean the same thing at all four. We test this on one
system, \sys, whose four interception points call one policy decision point (PDP) that cross-checks
two policy engines. We replay its test cases and \NgenCases{} generated cases through all four, and
measure how well its tests detect faults, using \NfeRuns{} injected front-end fault runs,
\NfeMutants{} mutants of front-end code and \NengMutants{} mutants of the engines. Wherever an
interception point could carry a case, its decision, rule verdicts and audit record equalled the SDK
path's (\NeqExtPerRun{} comparisons per run), except for one defect that the generated cases found:
for an integer argument of magnitude $2^{53}$ or more, the hook bridge never consulted the PDP, so the
call was refused with no decision record in fail-closed sessions and, in fail-open sessions, was
not stopped but only counted as an evidence gap. As the test adapters drive them,
\NneHookPctExt\% of drivable cases cannot be carried by the hook mode, and \NneMcpAdapter{} of the
\NneMcpAll{} for MCP reflect adapter configuration, not the gateway. The tests flagged all
\NfeApplicable{} applicable case-fault pairs, but \NdropArgsEvOnly{} of \NdropArgs{} dropped-argument
faults only by comparing the recorded request, not the decision. Of the engine mutants that loaded,
the two engines disagreed on some case for \NmSplitPct\% and \NmSurvPct\% survived every case. Median decision latency was
\NlatPSdk{}~ms (SDK), \NlatPHook{} (hook), \NlatPEgress{} (egress) and \NlatPMcp{} (MCP), measured on
one laptop.
\end{abstract}

\vspace{0.5em}

\clearpage

\section{Introduction}
\label{sec:intro}

\subsection{The problem}

AI agents act by calling tools: they search the web, read files, send network requests and run
commands. An agent can be steered by text it reads, so a hidden instruction in a web page or a
document may make it call a tool it should not, or call an allowed tool with dangerous arguments
\cite{greshake2023indirect,debenedetti2024agentdojo}.

A \emph{governance layer} reduces this risk. Before a tool call runs, the layer asks a \emph{policy
decision point} (PDP) whether the operator's policy allows the call. It then applies the answer and
records the decision as evidence. The operator writes the policy once.

Where the check happens is a deployment choice, and there are four common places:

\begin{itemize}
\item inside the agent's own code, as a wrapper around each tool that a software development kit
(SDK) provides (the \emph{SDK} mode);
\item in a gateway in front of a Model Context Protocol (MCP) server (the \emph{MCP} mode);
\item in a web (HTTP) proxy on the agent's network path, for agents whose code the operator does not
control (the \emph{egress} mode);
\item in a hook that a coding agent runs before each tool call (the \emph{hook} mode).
\end{itemize}

A single deployment often uses several of these at once, for defence in depth or because different
agents allow different hooks. The operator then relies on an unstated promise: the policy decides an
action in the same way, and leaves the same evidence, wherever the action was seen.

That promise can fail in three ways.

\begin{itemize}
\item \textbf{Translation.} The code at one interception point turns the action into a different
question for the decision point. It may lose an argument, rewrite a host name or change a number.
\item \textbf{Acting.} The code acts on the answer differently from the others. It may forward a
denied call, block an allowed one, or record nothing.
\item \textbf{Shared error.} The decision logic is wrong in a way that no single check can see,
because every check shares the error.
\end{itemize}

\subsection{A concrete example}
\label{sec:example}

Suppose an operator runs a research assistant that may search company filings with a tool called
\code{sec\_search}. The policy says that the argument \code{limit}, the number of results, must not
exceed 100. A call with \code{limit} set to $10^{30}$ must therefore be denied wherever it is seen,
and each denial should leave a record.

In our measurements, three interception points behave as expected. The SDK wrapper, the MCP gateway
and the egress proxy all send the call to the decision point, which denies it and records the
decision. The fourth, the hook for a coding agent, never asks the decision point. It cannot
serialise the number, treats the failure as ``input too large'' and stops. What happens next depends
on how the session is configured. The call is either refused with no decision record, or the hook
returns no verdict and only a counter of ``evidence gaps'' goes up.

The policy was correct and so was the decision point. One interception point simply stopped applying
the policy, and the only thing an agent had to do to trigger it was to write one large integer.
\Cref{sec:finding} gives the details.

\subsection{What this paper does}

We measure how consistently one system applies one policy across its four interception points. The
system is \sys, a runtime governance layer built by the author. It ships a conformance test suite,
and we run that suite against a live decision point. We send the suite's own test cases, and
\NgenCases{} further cases that we generated, through all four modes. We compare the decisions, the
reasons and the audit records that each mode produces.

We then ask how good the tests themselves are. We inject faults into each interception point, and we
make small changes (\emph{mutants}) to the source code of the interception points and of the policy
engines, and count how many of these the tests notice. Finally we measure the time each mode adds to
a decision.

\subsection{Research questions}

\begin{enumerate}[label=\textbf{RQ\arabic*.},leftmargin=3.4em,labelwidth=2.8em,itemsep=0.5em]
\item \textbf{Agreement.} When two interception modes can both carry an action, do they produce the
same decision, the same reasons and the same evidence?
\item \textbf{Coverage.} When a mode cannot carry what a policy needs, is that reported, and how
much of the policy does each mode actually cover?
\item \textbf{Oracle power.} How much can the tests detect: comparing modes, comparing the two policy
engines, and checking the expected result of each case? What do they miss by construction?
\item \textbf{Cost.} How much decision latency does each interception mode add?
\end{enumerate}

\subsection{Contributions}

\begin{enumerate}[label=\textbf{C\arabic*.},leftmargin=3.0em,labelwidth=2.4em,itemsep=0.5em]
\item \textbf{A four-mode differential measurement} of one decision point behind the four
interception points the system ships, on a live stack. Each case that a mode cannot carry is
reported as \emph{not expressible}, with a reason code, and the report separates limits of the
interception point from limits of the test adapters (\cref{sec:rq1,sec:rq2}).
\item \textbf{A generated grid of \NgenCases{} consistency-only cases} that found one divergence the
authored cases missed. The hook bridge mishandles integers of magnitude $2^{53}$ or more, which lets an
agent-chosen argument move a call onto the fail-open path (\cref{sec:finding}).
\item \textbf{An oracle-power study.} The study injects each of ten fault types into every call of one
interception point (\NfeRunsAll{} runs: \NfeRuns{} faults and a control). It also applies
\NfeMutants{} source mutants of the interception points and \NengMutants{} mutants of the policy engines
and the helpers they share. It shows which faults each test catches, that comparing decisions alone
misses translation faults which the evidence comparison catches, and two structural blind spots of
two-engine consensus (\cref{sec:rq3}).
\item \textbf{An open artifact:} the harness, the cases, the raw data and an independent re-check
of every number in the paper (\cref{sec:availability}).
\end{enumerate}

\subsection{Summary of findings}
\label{sec:summary}

Each finding states what its numbers mean. Details and caveats are in \cref{sec:results}.

\begin{enumerate}[label=\textbf{F\arabic*.},leftmargin=3.0em,labelwidth=2.4em,itemsep=0.55em]
\item \textbf{Agreement holds where it can be tested.} On every case that the MCP gateway, the
egress proxy or the hook bridge could carry, the decision, the ordered rule verdicts, the argument
rewrites, the risk level and the controls were identical to the SDK path's. That is \NeqExtPerRun{}
comparisons in each extended run, with no difference. The audit log held exactly one matching event
per call.
\item \textbf{One defect, found by generated cases.} For \NhookBigintCases{} generated cases, all
with an integer of magnitude $2^{53}$ or more, the hook bridge never consulted the decision point.
In a fail-closed session the call is refused without a decision record. In a fail-open session,
which includes enforce-mode roles left on the default fail mode, the call is not stopped and only an
evidence-gap counter changes.
\item \textbf{Large parts of the policy are not exercised outside the SDK path.} Of
\NextDrivable{} cases that the decision point can evaluate, \NneMcpAll{} cannot be carried by the MCP
mode, \NneEgressAll{} by the egress mode and \NneHookAll{} (\NneHookPctExt\%) by the hook mode. Most of the MCP
and hook gap, \NneMcpAdapter{} and \NneHookAdapter{} cases, comes from how the suite's adapters are
configured, not from the gateway or the bridge.
\item \textbf{The tests catch the faults they were built to catch, and the evidence check
matters.} The suite flagged all \NfeApplicable{} applicable pairs of case and injected fault. For
\NdropArgsEvOnly{} of \NdropArgs{} cases with dropped arguments, only the comparison of the recorded
request noticed, because the decision itself did not change.
\item \textbf{Two-engine consensus is strong but not complete.} For \NmSplitPct\% of loaded engine
mutants the two engines disagreed on at least one case, so the decision point would refuse to decide
for requests of that shape. Another \NmSurvPct\% of the loaded mutants survived every case. The consensus
check also misses two kinds of fault: \NmSharedExpOnly{} mutants in code shared by both engines were
caught only by a written expectation, and \NmAllMutOnly{} mutants changed only an argument rewrite,
which the shipped check does not compare.
\item \textbf{Latency differs by mode.} Median decision latency was \NlatPSdk{}~ms for the SDK
path, \NlatPHook{}~ms for the hook, \NlatPEgress{}~ms for the egress proxy and \NlatPMcp{}~ms for the
MCP gateway (measured on one laptop). The hook adds little to the SDK path; the MCP gateway adds the
most.
\end{enumerate}

\subsection{Scope of the claims}

The study covers one system, designed and built by the author, together with its test suite. The test
cases are synthetic requests against a purpose-built policy, not production traffic. The contribution
is the method and what it found. The paper does not claim that other governance layers behave in the
same way. Limitations are listed in \cref{sec:limitations}.

\subsection{Organisation of the paper}

\Cref{sec:background} defines the terms used in the paper. \Cref{sec:related} places the work among
related studies. \Cref{sec:system} describes the system, the property we test and the conformance
suite. \Cref{sec:method} explains the experiments and the test environment. \Cref{sec:results}
answers the four research questions, one subsection each, and includes the defect in the hook bridge.
\Cref{sec:discussion} discusses what the results mean and gives practical guidance.
\Cref{sec:limitations} lists limitations and future work, \cref{sec:availability} describes the data
and artifact, and \cref{sec:conclusion} concludes. Appendix~\ref{app:glossary} is a glossary of terms and
abbreviations.

\section{Background and terminology}
\label{sec:background}

This section explains, in plain words, the pieces of a governance layer and the testing terms used
later. Every term is also listed in the glossary (Appendix~\ref{app:glossary}).

\subsection{AI agents, tools and tool calls}

An \emph{AI agent} is a program in which a large language model chooses what to do next. The agent
acts on the world through \emph{tools}: functions or services such as a web search, a file reader or
a shell. A \emph{tool call} names a tool and gives it \emph{arguments}, for example the search text
and a limit on the number of results. Some calls also have a \emph{target}, which is a destination
such as a web address, a network address or a file path.

Text that the agent reads can contain instructions that steer the model, which is called prompt
injection \cite{greshake2023indirect}. The model's choice of tool, arguments and target therefore
cannot be trusted on its own \cite{debenedetti2024agentdojo}. Operators add a check before the call
runs.

\subsection{Governance layers, policy and the decision point}

A \emph{governance layer} is software that checks each tool call before it runs. The check is made by
a \emph{policy decision point} (PDP). The PDP is a service that receives a description of the
proposed call, called the \emph{decision input}, and returns a decision. The decision input names
the organisation, the \emph{role} of the caller (a named set of permissions, such as ``analyst''), the
caller's identity, the tool, the arguments, the target and a few counters about recent calls.

The operator writes a \emph{policy}: for each role, a set of rules. The PDP evaluates every rule and
reports one verdict per rule: pass, fail, or not applicable. The answer therefore lists all reasons,
not only the first failure. The decision itself is ALLOW or DENY. The PDP may also return
\emph{argument rewrites}, which are changes the caller must apply to the arguments, such as removing an
access key before the call is sent.

The rules fall into six groups, listed in \cref{tab:groups}.

\begin{table}[htbp]
\centering\tablesetup
\caption{The six rule groups of the policy.}
\label{tab:groups}
\begin{tabular}{@{}l L{4.4cm} L{8.6cm}@{}}
\toprule
Group & What it checks & Example \\
\midrule
identity & who is calling and how & only token and OpenID Connect (OIDC) authentication allowed; credential not expired \\
authorization & which tools a role may use & the role may call \code{sec\_search} and \code{doc\_fetch}, nothing else \\
arguments & limits on argument values & \code{limit} at most 100; query must not match a blocked pattern \\
target & where the call goes & destination host on an allow-list; address outside a denied range; blocked file path \\
quota & how often & at most 3 calls per minute; three identical calls in five minutes count as a loop \\
authority & on whose behalf & delegation at most one hop deep; a delegated call must name the grant that authorises it \\
\bottomrule
\end{tabular}
\end{table}

\subsection{Observe and enforce, fail-closed and fail-open}
\label{sec:modes}

Each role runs in one of two \emph{modes}. In \emph{observe} mode, the PDP evaluates the policy and
records the decision, but the action always proceeds. Operators use this to roll out a policy without
breaking work. In \emph{enforce} mode, the action proceeds only if the decision is ALLOW. The PDP
reads the mode from the policy and ignores any mode that the caller claims.

A role also has a \emph{fail mode}, which says what happens when no usable decision is available, for
example when the PDP cannot be reached. With fail mode \code{closed}, the call is refused
(\emph{fail-closed}). With fail mode \code{open\_with\_alert}, the call proceeds and an alert is
raised (\emph{fail-open}). In the system studied here the default fail mode is \code{open\_with\_alert}.
An enforce-mode role is therefore not fail-closed unless its fail mode is set to \code{closed}.

\subsection{Audit evidence}
\label{sec:evidence}

The PDP seals one \emph{audit event} for each decision before it answers. The events of an
organisation form a \emph{hash chain}: each event contains the hash of the previous one, so a change
to an earlier event can be detected. A call for which no decision event exists is an \emph{evidence
gap}. Claims such as ``this call was allowed under policy version 12'' are only as good as these
records. For this reason the tests in this paper compare the records, and not only the decisions.

\subsection{Interception points, modes and front ends}
\label{sec:interception}

\begin{figure}[htbp]
\centering
\resizebox{\linewidth}{!}{%
\begin{tikzpicture}[font=\small,
  box/.style={draw,rounded corners=2pt,align=center,fill=white,inner sep=4pt,minimum height=9mm},
  small/.style={font=\footnotesize,align=center},
  arr/.style={-{Stealth[length=2.2mm]},thick}]
  \node[box,minimum width=2.6cm,minimum height=1.8cm] (agent) at (0,0) {AI agent\\[-1pt]{\footnotesize model and tools}};
  \node[box,minimum width=5.2cm,fill=black!4] (sdk) at (5.6,2.55) {\textbf{SDK wrapper}\\[-1pt]{\footnotesize in the agent's own process}};
  \node[box,minimum width=5.2cm,fill=black!4] (mcp) at (5.6,0.85) {\textbf{MCP gateway}\\[-1pt]{\footnotesize between agent and MCP server}};
  \node[box,minimum width=5.2cm,fill=black!4] (egr) at (5.6,-0.85) {\textbf{Egress proxy}\\[-1pt]{\footnotesize on the network path}};
  \node[box,minimum width=5.2cm,fill=black!4] (hook) at (5.6,-2.55) {\textbf{Hook bridge}\\[-1pt]{\footnotesize in a coding agent's pre-tool hook}};
  \begin{scope}[on background layer]
    \node[draw,dashed,rounded corners=3pt,fit=(sdk)(mcp)(egr)(hook),inner sep=6pt,
      label={[small,anchor=south]north:Four interception points (front ends)}] (fe) {};
  \end{scope}
  \node[draw,rounded corners=3pt,minimum width=5.4cm,minimum height=5.6cm,anchor=center] (pdp) at (14.6,0) {};
  \node[small] at (14.6,2.45) {\textbf{Policy decision point (PDP)}};
  \node[box,minimum width=4.2cm] (py) at (14.6,1.0) {Python engine};
  \node[box,minimum width=4.2cm] (opa) at (14.6,-0.5) {OPA engine (Rego)};
  \node[small] at (14.6,-2.0) {answers only if\\both engines agree};
  \node[box,minimum width=5.4cm,fill=black!4] (audit) at (14.6,-5.0) {\textbf{Audit log}\\[-1pt]{\footnotesize hash-chained, one event per decision}};
  \node[box,minimum width=5.4cm,fill=black!4] (tool) at (5.6,-5.6) {\textbf{Protected tool, server or destination}};
  \draw[arr] (agent.east) -- node[above,small]{tool call} (fe.west |- agent.east);
  \draw[arr] ([yshift=4mm]fe.east |- pdp.west) -- node[above,small]{decision input} ([yshift=4mm]pdp.west);
  \draw[arr] ([yshift=-4mm]pdp.west) -- node[below,small]{decision, reasons} ([yshift=-4mm]fe.east |- pdp.west);
  \draw[arr] (pdp.south) -- node[right,small]{seals one event} (audit.north);
  \draw[arr] (fe.south) -- node[right,small,pos=0.5]{forwarded only if allowed} (tool.north);
\end{tikzpicture}}
\caption{The studied system. The agent's action reaches one of four front ends. Each front end asks
the same decision point, applies the answer, and forwards the call only if it is allowed. The decision
point evaluates the policy with two engines and seals an audit event for every decision.}
\label{fig:architecture}
\end{figure}
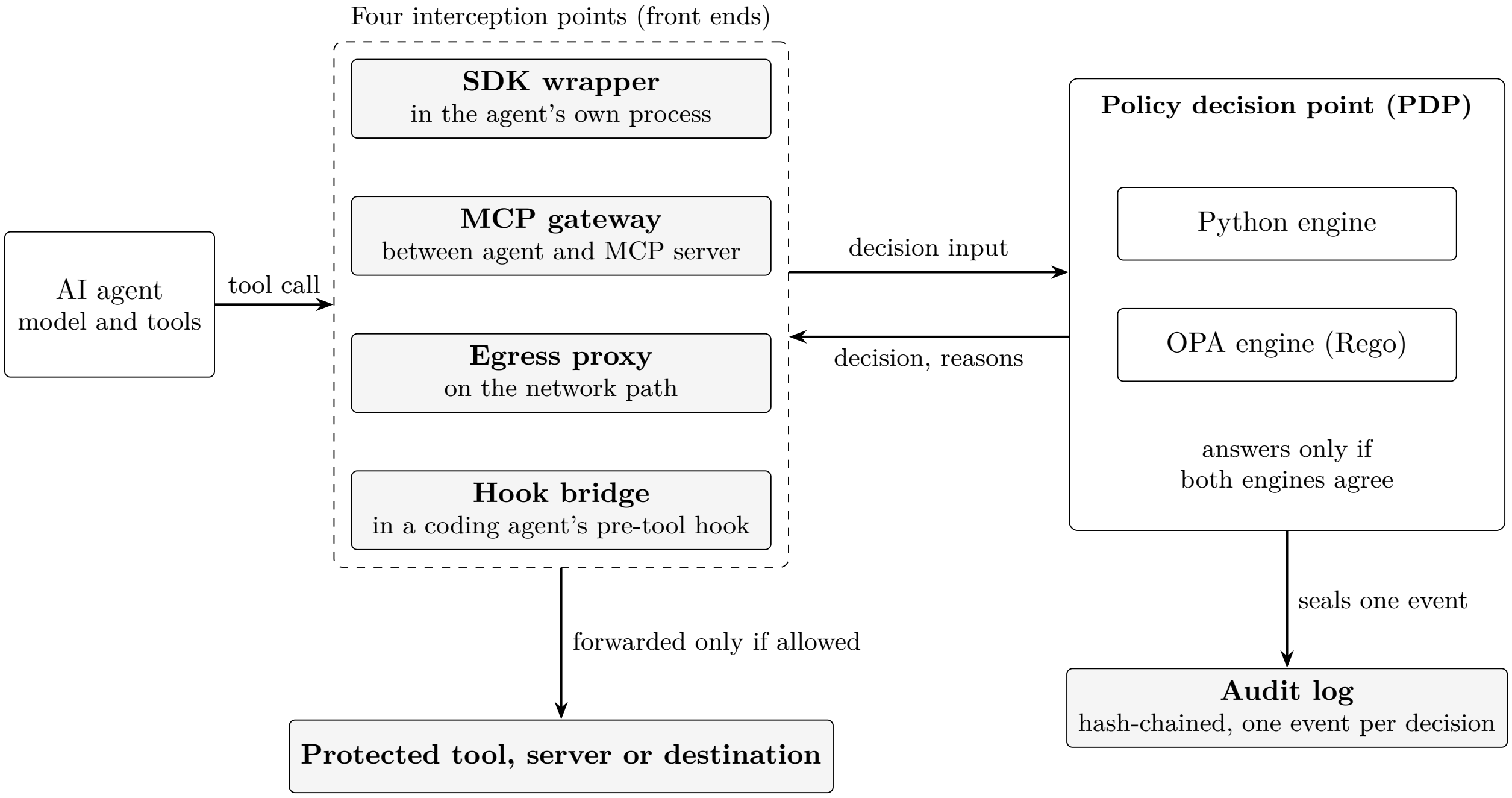

An \emph{interception point} is a place where an action can be checked. The code that sits there and
talks to the PDP is a \emph{front end}. Its configuration is an \emph{interception mode}. The studied
system has four, described in \cref{tab:modes} and drawn in \cref{fig:architecture}. In this paper
\emph{native modes} are the three whose front ends are real components of the system (MCP, egress,
hook). The SDK mode is the baseline.

\begin{table}[htbp]
\centering\tablesetup
\caption{The four interception modes. The identifier in parentheses is the name used in the data files.}
\label{tab:modes}
\begin{tabular}{@{}L{2.7cm} L{6.2cm} L{6.9cm}@{}}
\toprule
Mode & Where it sits & What it can carry \\
\midrule
SDK (\code{sdk}) & A wrapper around each tool function, in the agent's own process. & Every argument and the tool definition; no network target. \\
MCP (\code{mcp}) & A gateway between an MCP client and one MCP server \cite{mcp-spec-2025-06-18}. & Tool name and arguments from \code{tools/call}; the client's identity comes from the gateway's configuration. \\
Egress (\code{egress}) & An HTTP(S) proxy on the agent's network path. & The destination of the request and the top-level fields of a JSON (JavaScript Object Notation) body as arguments. \\
Hook (\code{hook}) & A local bridge that answers a coding agent's pre-tool hook \cite{claudecode-hooks}. & The tool name and input that the coding agent reports; a file path or web address for the agent's own file and fetch tools. \\
\bottomrule
\end{tabular}
\end{table}

\begin{shows}
The four modes differ in where they sit and in what they can see, not in the policy they apply. The
SDK wrapper sees the arguments but no network target. The egress proxy sees a destination and a body
but no tool schema. These differences decide which parts of a policy each mode can enforce, and
\cref{sec:rq2} measures them.
\end{shows}

\subsection{Two engines and the consensus check}
\label{sec:engines}

The PDP evaluates each request with two independent implementations of the policy. One is a Python
\emph{reference engine}. The other is OPA, the Open Policy Agent \cite{opa-rest-api}, running a
hand-written module in the Rego language over data compiled from the same policy file. If the two
engines disagree on the set of failing rules, the PDP answers HTTP status 503 and returns no
decision. A disagreement is treated as a fault and is not resolved by preferring one engine. This
is N-version programming \cite{avizienis1985nversion} used as a check at run time. We call it
\emph{engine consensus}. Argument rewrites come from the Python engine only.

Some code is shared by construction, to make agreement reachable. Regular expressions are compiled
into one dialect, address lists are expanded in advance, and paths and argument paths are parsed
in advance, all by Python \emph{helpers} that feed both the Python engine and the data that OPA
reads. A defect in a shared helper changes both engines in the same way. We call this a
\emph{common-mode fault}. Consensus cannot see it.

\subsection{Testing terms}

\emph{Differential testing} \cite{mckeeman1998differential} gives the same input to systems that
should behave alike and compares their outputs. Here the same case goes through four modes. An
\emph{oracle} is the rule that decides whether an observed result is correct. The tests in this paper
use three oracles: comparison across modes, comparison across the two engines, and the
\emph{expectation} written into a case.

\emph{Fault injection} breaks a component on purpose to see whether the tests notice.
\emph{Mutation testing} \cite{jia2011mutation} makes small changes to source code. Each changed copy
is a \emph{mutant}. A mutant is \emph{killed} if some test result differs because of the change. A
mutant that no test notices \emph{survives}. Some survivors are \emph{equivalent mutants}: changes
that do not alter behaviour at all. The survival rate is therefore an upper bound on undetected
defects. A mutant that cannot be loaded is \emph{stillborn}. A \emph{first-order} mutant contains
exactly one change.

\section{Related work}
\label{sec:related}

We group related studies by theme. Each group ends with how this paper differs. A search of arXiv and
the web in November 2026 found no earlier measurement that compares all interception points of one
deployed governance system. The search was not exhaustive: it did not cover bibliographic databases
or venue proceedings.

\subsection{Coverage and consistency of runtime controls}

guardscope \cite{toyoda2026guardscope} predicts which of 15 action paths a runtime control can
check. It classifies executed actions as checked, bypassed or prevented, using kernel-level ground
truth. It answers whether a control is on the path of an action. Cross-Interface Authorization
\cite{zhu2026crossinterface} is a protocol that turns requests from five interfaces of computer-use
agents into canonical effects. It comes with proofs and with an independent checker that rejects
mutated normalisers. GatewayBench \cite{zheng2026gatewaybench} compares seven MCP gateways on task
completion, tool-use efficiency and resistance to attacks.

Our work differs in what it assumes and what it asks. We assume that a control is on the path and ask
whether controls on different paths decide alike. Cross-Interface Authorization designs for equivalence; we measure an
existing system and account for what each mode cannot express. GatewayBench does not compare
gateways with other interception points.

\subsection{Surface form and parsing}

ToolRosetta \cite{demirkiran2026toolrosetta} presents the same action as a code call or as a shell
command. It finds that four of five specialised guards show false-positive rates that are 21 to 47
points higher on one benchmark under the shell form, while frontier guards stay stable. Patodiya
\cite{patodiya2026postwire} applies differential testing to the way four Python agent integrations
transform MCP tool results after they leave the wire.

Our generated value grid belongs to the same family of tests. The divergence we found is a
serialisation pre-check in one front end, and not a parser difference.

\subsection{Differential and mutation testing of policy and authorization engines}

Differential testing compares comparable systems on generated inputs
\cite{mckeeman1998differential}, and Csmith made it a standard tool for compilers
\cite{yang2011csmith}. The Cedar authorization language checks its Rust authorizer against Lean
models with differential random testing \cite{disselkoen2024howwebuiltcedar,cutler2024cedar}.
Mutation analysis \cite{jia2011mutation} has been applied to access-control policies
\cite{martin2007fault}. Metamorphic testing \cite{chen2018metamorphic} relates the outputs of
related inputs. The observe/enforce comparison in the suite is such a relation.

This paper applies differential testing across the front ends of one deployed system. In that system,
both engines also run inside the decision point, and the decision point fails closed on disagreement.
This turns the test oracle into a run-time control. We mutate the engines and the front ends, not the
policy.

\subsection{Enforcement points}

Complete mediation \cite{saltzer1975protection}, the requirement that every access be checked, and
the reference monitor concept \cite{anderson1972planning} explain why several interception points
exist. An industrial gateway for large language model traffic evaluates policy with OPA at the
network edge \cite{sasidharan2026cosmos}. A harness for coding agents adds a reference monitor with
capability tokens and taint tracking \cite{hegelich2026activesecharness}.

None of these works reports whether the decisions of different interception points agree. That is the
question we measure.

\section{The property and the system studied}
\label{sec:system}

\subsection{The system under test}
\label{sec:sut}

The system under test is \sys{} version 0.1.0-dev \cite{oathpostrepo}, a runtime governance layer for
AI agents built by the author. It has the following parts, all of which take part in the
measurements.

\begin{itemize}
\item \textbf{The SDK}, a Python library. It wraps a tool function, asks the PDP whether a call may
proceed, applies the answer and records the outcome.
\item \textbf{The policy decision point (PDP)}, an HTTP service. It receives a decision input,
evaluates the policy of the caller's role with two engines (\cref{sec:engines}), seals an audit event
and returns the decision. The three native front ends call its endpoint \code{/v1/decide/consume},
which also spends a human approval if one is required. The SDK path calls \code{/v1/decide}.
\item \textbf{The policy}, a file in the YAML text format. A compiler turns it into a policy bundle that both engines
evaluate; OPA reads the bundle as data.
\item \textbf{Four front ends}: the SDK wrapper, the MCP gateway, the egress proxy and the hook
bridge (\cref{sec:interception}).
\item \textbf{The audit log}, a hash-chained table in PostgreSQL (\cref{sec:evidence}).
\item \textbf{A conformance suite}, shipped with the system and described in \cref{sec:suite}.
\end{itemize}

\Cref{fig:listing} shows a decision input and the answer for the call of \cref{sec:example}, with
most fields left out.

\begin{figure}[htbp]
\centering
\begin{tcolorbox}[enhanced,colback=black!3,colframe=black!40,boxrule=0.4pt,arc=1pt,left=6pt,right=6pt,top=4pt,bottom=4pt]
\begin{minipage}[t]{0.50\linewidth}
\begin{Verbatim}[fontsize=\scriptsize]
decision input
{ "role": "analyst__enforce",
  "identity": {
    "principal": "agent:demo",
    "auth_method": "token" },
  "tool": { "name": "sec_search" },
  "args": { "scope": "10-K",
            "limit": 10**30 },
  "target": { },
  "context": {
    "interception_mode": "sdk" } }
\end{Verbatim}
\end{minipage}\hfill
\begin{minipage}[t]{0.47\linewidth}
\begin{Verbatim}[fontsize=\scriptsize]
answer
{ "decision": "DENY",
  "effective_decision": "DENY",
  "reasons": [
    { "rule_id": "arguments.max_value",
      "verdict": "fail" },
    ... one entry for every other rule ... ],
  "mutations": [],
  "risk_level": "MEDIUM",
  "audit_event_id": "..." }
\end{Verbatim}
\end{minipage}
\end{tcolorbox}
\caption{An abridged decision input (left) and answer (right). The role name ends in \code{\_\_enforce}
because the suite creates an enforce and an observe twin of every role. The number \code{10**30} stands
for the integer $10^{30}$. The field \code{mutations} lists the argument rewrites.}
\label{fig:listing}
\end{figure}

\subsection{The property: one policy, one meaning}
\label{sec:property}

Let $X$ be the set of cases that the PDP can evaluate. For a mode $m$, let $E_m \subseteq X$ be the
cases that mode $m$ can \emph{express}: the cases that its front end can present to the PDP without
changing what they ask. For a case $c \in E_m$, let $\mathit{out}_m(c)$ be the \emph{outcome} recorded
when $c$ is sent through mode $m$. The outcome has eight fields: the decision, the effective decision
(what actually happened to the action), the ordered verdicts of all rules, the argument rewrites,
the obligations, the risk level, the satisfied controls and the policy version.

\begin{statement}{Property P1: same answer}
For every native mode $m \in \{\text{MCP},\ \text{egress},\ \text{hook}\}$ and every case $c \in E_m$:
\begin{equation}
\mathit{out}_m(c) \;=\; \mathit{out}_{\text{SDK}}(c) .
\label{eq:equivalence}
\end{equation}
\end{statement}

In words: whenever a mode can carry a case, it must produce the same outcome as the SDK path. The
property says nothing about cases outside $E_m$. They are reported separately
(\cref{sec:notexpressible}).

\begin{statement}{Property P2: same evidence}
For every call through a native mode, the audit chain holds exactly one decision event. The event
records what the front end sent to the PDP and what the PDP answered. The front end acted on that
answer: it ran the protected tool only on an effective ALLOW.
\end{statement}

\subsection{The conformance suite}
\label{sec:suite}

The suite is a Python package that ships with the system. It replays \emph{cases} against a live
PDP. A case is a recorded decision input and, optionally, an \emph{expectation}: the decision and the
failing rules that its author intended. Cases that carry no expectation assert only consistency:
whatever the decision is, all modes and both engines must agree.

An \emph{adapter} presents a case to the PDP through one mode and returns the outcome. There is one
adapter per mode.

\begin{itemize}
\item The \textbf{SDK adapter} posts the request that the SDK would build. This is the baseline.
\item The \textbf{MCP adapter} drives the real MCP gateway against a stub MCP server.
\item The \textbf{egress adapter} sends an HTTP request through the real egress proxy to a stub target
server.
\item The \textbf{hook adapter} feeds hook-shaped JSON to the real hook bridge. No coding agent is
involved.
\end{itemize}

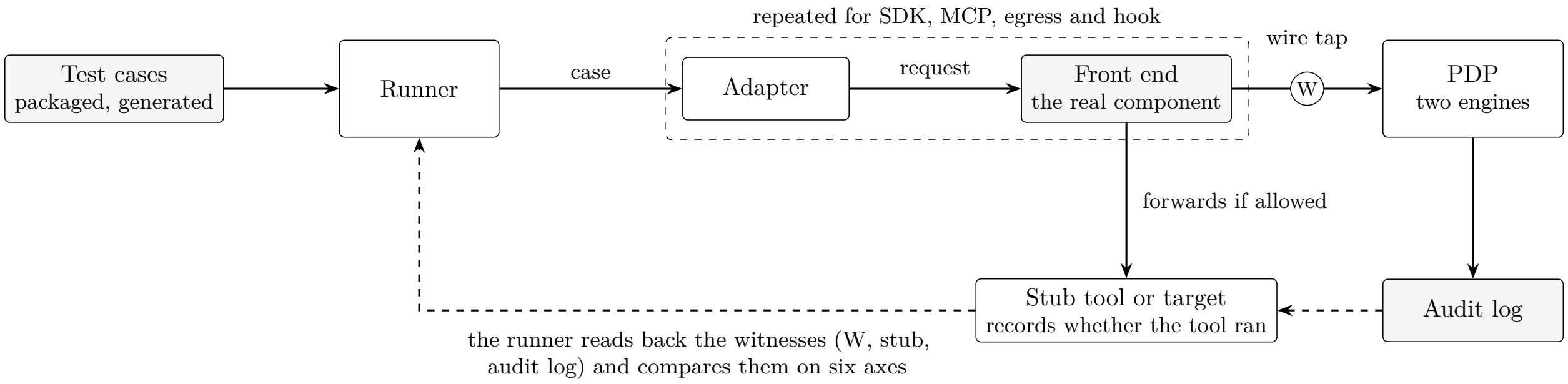
\begin{figure}[htbp]
\centering
\resizebox{\linewidth}{!}{%
\begin{tikzpicture}[font=\small,
  box/.style={draw,rounded corners=2pt,align=center,fill=white,inner sep=4pt,minimum height=9mm},
  small/.style={font=\footnotesize,align=center},
  arr/.style={-{Stealth[length=2.2mm]},thick},
  back/.style={-{Stealth[length=2.2mm]},thick,dashed}]
  \node[box,minimum width=2.7cm,fill=black!4] (cases) at (0,0) {Test cases\\[-1pt]{\footnotesize packaged, generated}};
  \node[box,minimum width=2.3cm,minimum height=1.4cm] (runner) at (4.4,0) {Runner};
  \node[box,minimum width=2.4cm] (adapter) at (9.4,0) {Adapter};
  \node[box,minimum width=3.0cm,fill=black!4] (fe) at (14.6,0) {Front end\\[-1pt]{\footnotesize the real component}};
  \node[box,minimum width=2.6cm,minimum height=1.4cm] (pdp) at (19.6,0) {PDP\\[-1pt]{\footnotesize two engines}};
  \node[box,minimum width=2.6cm,fill=black!4] (audit) at (19.6,-3.2) {Audit log};
  \node[box,minimum width=3.0cm] (stub) at (14.6,-3.2) {Stub tool or target\\[-1pt]{\footnotesize records whether the tool ran}};
  \begin{scope}[on background layer]
    \node[draw,dashed,rounded corners=3pt,fit=(adapter)(fe),inner sep=7pt,
      label={[small,anchor=south]north:repeated for SDK, MCP, egress and hook}] {};
  \end{scope}
  \draw[arr] (cases) -- (runner);
  \draw[arr] (runner) -- node[above,small]{case} (adapter);
  \draw[arr] (adapter) -- node[above,small]{request} (fe);
  \draw[arr] (fe) -- (pdp);
  \node[draw,circle,inner sep=1.2pt,fill=white,font=\scriptsize] (tap) at ($(fe.east)!0.5!(pdp.west)$) {W};
  \node[small,anchor=south] at ([yshift=2mm]tap.north) {wire tap};
  \draw[arr] (pdp) -- (audit);
  \draw[arr] (fe) -- (stub);
  \node[small,anchor=west] at ([xshift=1mm]$(fe.south)!0.5!(stub.north)$) {forwards if allowed};
  \draw[back] (audit.west) -- (stub.east);
  \draw[back] (stub.west) -- (runner.south |- stub.west) -- (runner.south);
  \node[small,below=1.5mm,text width=7.6cm] at ($(stub.west)!0.5!(runner.south |- stub.west)$)
    {the runner reads back the witnesses (W, stub, audit log) and compares them on six axes};
\end{tikzpicture}}
\caption{The measurement set-up. The runner sends every case through each adapter. Each adapter drives
the real front end of its mode, which asks the PDP. A wire tap on every front-end-to-PDP leg and a stub
upstream behind the front end act as independent witnesses. The runner reads the audit log back and
compares all witnesses.}
\label{fig:testsetup}
\end{figure}

The adapters do not trust the front end's own account of what it did. A recording transport on the
leg between the front end and the PDP (the \emph{wire tap}) captures the exact question and answer. A
stub upstream or target records whether the protected tool actually ran. The suite then compares
these witnesses with what the client experienced. \Cref{fig:testsetup} shows the arrangement.

The runner compares outcomes on six \emph{axes}, listed in \cref{tab:axes}. The overall result of a run
is called the \emph{verdict}. It is \code{CONFORMANT} if every axis passed and every required mode was
exercised, \code{DIVERGENT} if any axis found a difference, \code{INCOMPLETE} if nothing diverged but
some mode was not measured, and \code{ERROR} if the suite could not reach a conclusion.

\begin{table}[htbp]
\centering\tablesetup
\caption{The six comparison axes of the conformance suite.}
\label{tab:axes}
\begin{tabular}{@{}l L{12.2cm}@{}}
\toprule
Axis & Question it answers \\
\midrule
Observe/enforce & Do the two twins of a role, identical except for their mode, give the same decision and differ only in whether the action proceeds? \\
Cross-mode & Does each native mode produce the same outcome as the SDK mode (Property P1, same answer)? \\
Engine consensus & Do the Python engine and OPA agree on the failing rules, the decision and the argument rewrites, and does the PDP add only declared reasons? \\
Expectation & Does each case do what its author wrote it to do? \\
Evidence & Did each front end act on its verdict, send the PDP what the caller supplied, and leave exactly one agreeing event in the audit chain (Property P2, same evidence)? \\
Visibility & Does an MCP tool listing show exactly the tools that a call would not refuse as unauthorised? \\
\bottomrule
\end{tabular}
\end{table}

\subsection{Cases that a mode cannot carry}
\label{sec:notexpressible}

A mode cannot always present a case faithfully. An HTTP request, for example, has no field for a
delegation chain. The adapter therefore decides whether it can carry a case before it sends any
request. If it cannot, the suite reports the case as \emph{not expressible} for that mode, with a
stable \emph{reason code}. Such a case is never counted as equal. An unmeasured mode keeps the verdict
\code{INCOMPLETE} unless it is waived by name.

\Cref{tab:reasons} defines the reason codes. The last column separates two kinds of limit. A
\emph{front-end limit} is a property of the shipped front end. An \emph{adapter limit} is a property of
how the suite's adapter is configured, so a different adapter could carry the case through the same
front end.

\begin{table}[htbp]
\centering\tablesetup
\caption{Reason codes for \emph{not expressible}, and the kind of limit each one is.}
\label{tab:reasons}
\begin{tabular}{@{}l L{7.6cm} L{3.4cm}@{}}
\toprule
Code & Meaning & Kind of limit \\
\midrule
\code{auth\_method\_fixed} & The front end authenticates callers by one method that is fixed per deployment. A case that asks for another method cannot be carried. & front end \\
\code{no\_credential\_expiry} & The front end carries no credential-expiry time. & front end \\
\code{no\_delegation\_chain} & The case names a delegation chain that the mode's requests do not carry. An HTTP request and a hook call have none. The suite's MCP adapter sends none, although the gateway can read one from the request's \code{\_meta} field. & front end (egress, hook); adapter (MCP) \\
\code{no\_network\_target} & The case has a network target. The suite's MCP adapter configures no per-tool target rules. The hook adapter presents every tool as an MCP tool and not as the coding agent's own file and fetch tools, from which the bridge derives a path or host. & adapter \\
\code{host\_not\_canonical} & The proxy presents only a canonical host: lower case, no port, no trailing dot, no zone identifier. The case spells the host differently. & front end \\
\code{host\_unconnectable} & The host is an ambiguous numeric spelling (such as \code{127.1} or \code{0x7f000001}) that the proxy neither resolves nor dials. & front end \\
\code{path\_not\_expressible} & The target path cannot be carried verbatim in an HTTP request path. & front end \\
\code{target\_address\_claim} & The case claims a resolved address or address range, which the proxy pins itself. & front end \\
\code{target\_always\_present} & An HTTP request always has a destination. A case with no target on a role that restricts destinations cannot be presented. & front end \\
\code{tool\_name\_invalid} & The tool name is empty or longer than 256 characters, which the protocol does not allow. & front end \\
\bottomrule
\end{tabular}
\end{table}

We call a case \emph{adapter-limited} for a mode when every reason code on it is an adapter limit:
\code{no\_network\_target} for MCP and hook, and \code{no\_delegation\_chain} for MCP.

\subsection{What counts as a failure, and the threat model}
\label{sec:threat}

The agent may be prompt-injected, so it may choose any tool name, argument value, target or path that
its interception mode lets it express. The governance components are honest but may contain defects.
The defects that we inject are accidental and are not the work of an adaptive adversary. Two outcomes
count as failures.

\begin{statement}{Failure A: policy bypass}
An action runs that the policy denies.
\end{statement}

\begin{statement}{Failure B: unrecorded action}
An action runs, or is refused, without a decision record that says so.
\end{statement}

A refusal with a record is acceptable even when another mode would have allowed the call. We report it
as a divergence but not as a bypass. Out of scope are an agent that avoids every interception point
(measured by guardscope \cite{toyoda2026guardscope}), a compromised PDP or database, and the side
effects of a shell command after a hook has allowed it.

\section{Experimental method}
\label{sec:method}

\subsection{Overview}

Five experiments answer the four research questions. \Cref{tab:experiments} lists them. The first
two research questions use the same runs. Each experiment is described in its own subsection below,
after the test environment and the case sets.

\begin{table}[htbp]
\centering\tablesetup
\caption{The experiments, what they vary, what they measure and how often each was run.}
\label{tab:experiments}
\begin{tabular}{@{}L{3.0cm} L{2.4cm} L{3.4cm} L{3.4cm} L{2.4cm}@{}}
\toprule
Experiment & Answers & What varies & What is measured & Runs \\
\midrule
Four-mode runs (\cref{sec:m-fourmode}) & RQ1 agreement, RQ2 coverage & case set; mode & equality of outcomes; cases not expressible & packaged \NpkgRuns{}, extended \NextRuns{} \\
Injected front-end faults (\cref{sec:m-faults}) & RQ3 oracle power & fault type; mode & cases flagged, and by which axis & one per fault and mode (\NfeRuns{}) plus a control \\
Front-end source mutants (\cref{sec:m-fesource}) & RQ3 oracle power & one source change per run & killed or survived, and by which axis & one per mutant (\NfeMutants{}) \\
Engine mutants (\cref{sec:m-engine}) & RQ3 oracle power & one source change per run & killed or survived, and by which signal & one per mutant (\NengMutants{}) \\
Latency (\cref{sec:m-latency}) & RQ4 cost & mode & time per decision & \NlatRuns{} runs of \NlatCallsPerRun{} calls \\
\bottomrule
\end{tabular}
\end{table}

\subsection{Test environment}
\label{sec:environment}

All experiments ran on one laptop. \Cref{tab:environment} lists the hardware and the software. The
decision point, the front ends, the suite and the timing client ran on this machine and talked over
the loopback network interface. Each run started its own OPA container, PostgreSQL database and Redis
cache.

\begin{table}[htbp]
\centering\tablesetup
\caption{The test environment.}
\label{tab:environment}
\begin{tabular}{@{}l L{11.5cm}@{}}
\toprule
Item & Value \\
\midrule
Machine & Laptop, Windows 11 (build 10.0.26300) \\
CPU & Intel Core i9-14900HX \\
Memory & 16 GB RAM \\
Python & 3.13.13 \\
System under test & \sys{} 0.1.0-dev: one PDP process with one worker, the SDK, the MCP gateway, the egress proxy, the hook bridge and the conformance suite \\
Policy engine & OPA 1.22.2 in a Docker container (image identifier recorded in the data) \\
Database & PostgreSQL 16 in a TimescaleDB Docker image (referenced by tag) \\
Cache & Redis 7 in a Docker image (referenced by tag) \\
Container runtime & Docker Desktop \\
\bottomrule
\end{tabular}
\end{table}

The latency runs were taken while no other test process was running, which was checked before and
after each run.

\subsection{Case sets}
\label{sec:m-cases}

\subsubsection{Packaged set}
The packaged set is the suite's own: \NpkgCases{} cases and \NpkgMalformed{} malformed requests. Of the
cases, \NpkgDrivable{} can be driven through the PDP. The other \NengineOnly{} need token or cost
spending that only the PDP computes, and run against the engines only. The set contains a failing
case for \NpkgFailRules{} distinct rules.

\subsubsection{Extended set}
The extended set is the packaged set plus \NgenCases{} generated cases, \NextCases{} in all. The
generated cases carry no expectation: whatever the decision is, all modes and both engines must agree.

The generator crosses each rule group with input shapes at which JSON implementations are known to
differ. \Cref{tab:gen} lists the families. Examples are integers beyond $2^{53}$, negative zero,
values near the largest double, booleans where numbers are expected, numeric strings, Unicode line
separators, combining marks, full-width and look-alike characters, and deep nesting. The target
families use IPv4 and IPv6 address spellings (mapped, NAT64, 6to4, compatible) against four network
policies, and 32 path spellings. All hosts are IP literals, never names that need the Domain Name
System (DNS), so no case depends on a resolver.

\begin{table}[htbp]
\centering\tablesetup
\caption{The \NgenCases{} generated cases, by rule group and family. Each family varies one dimension
of the input.}
\label{tab:gen}
\begin{tabular}{@{}llrL{0.5\textwidth}@{}}
\toprule
Rule group & Family & Cases & What varies \\
\midrule
arguments & arg-code & 11 & ASCII digits vs other digits \\
 & arg-depth & 13 & JSON equality: 0 vs 0.0 vs false vs "0" \\
 & arg-docpath & 25 & traversal spellings and length \\
 & arg-limit & 37 & numeric bound: integers to $\pm 2^{63}$, $10^{30}$, floats, $-0.0$, types \\
 & arg-misc & 28 & other constraint types, special and odd keys, deep and wide values \\
 & arg-options & 16 & nested keys and blocked nested fields \\
 & arg-query & 33 & length and regex: Unicode, separators, homoglyphs, combining marks \\
 & arg-scope & 13 & allowed values: case, hyphen look-alikes, types \\
 & arg-ticker & 16 & anchored regex: newlines, full-width, accents, types \\
 & arg-urls & 11 & wildcard list elements \\
\addlinespace[2pt]
authority & authority-grid & 324 & role $\times$ auth method $\times$ depth $\times$ source $\times$ expiry \\
\addlinespace[2pt]
authorization & toolname & 25 & tool-name spellings \\
\addlinespace[2pt]
quota & quota-grid & 24 & rate and loop limits $\times$ 0--5 prior calls, same/different \\
\addlinespace[2pt]
target & target-grid & 720 & 4 network policies $\times$ 30 IPv4/IPv6 spellings $\times$ 6 paths \\
 & target-paths & 64 & 32 path spellings $\times$ 2 hosts \\
\midrule
Total & & 1,360 & \\
\bottomrule
\end{tabular}
\end{table}

\subsection{Four-mode runs}
\label{sec:m-fourmode}

Each run starts a private OPA container, a private PostgreSQL database and Redis cache, and a fresh
PDP process that serves the case set's policy under a fresh organisation. It then runs the suite's
command line with all four native modes and reads the audit chain back.

The suite's adapters are registered through its option for additional adapters, inside a wrapper that
only times each call. The wrapper also registers three \emph{label-only adapters}. They send the SDK
request unchanged except for the value of the interception-mode field. They test whether the PDP's
answer depends on that label.

We ran the packaged set \NpkgRuns{} times and the extended set \NextRuns{} times. The outcomes were
identical in every run.

\subsection{Injected front-end faults}
\label{sec:m-faults}

The suite has a self-test that injects each of ten fault types into one case. We apply each fault to
\emph{every} call of one front end, over the whole packaged set, one fault per run, on the live
stack. We do this for each of the three native modes, which gives \NfeRuns{} runs, and add a clean
control run (\NfeRunsAll{} runs in all). \Cref{tab:faultdefs} defines the faults. Four are \emph{wire
faults}, which corrupt the exchange between the front end and the PDP or the front end's action on
the answer. Six are \emph{outcome faults}, which corrupt the outcome that the front end reports.

\begin{table}[htbp]
\centering\tablesetup
\caption{The ten injected faults. The identifier in parentheses is the name used in the data files.}
\label{tab:faultdefs}
\begin{tabular}{@{}l L{5.2cm} L{7.8cm}@{}}
\toprule
Kind & Fault & What it does \\
\midrule
Wire & Forward a denied call (\code{forward\_denied}) & The front end forwards a call that the PDP denied in enforce mode. \\
Wire & Refuse an allowed call (\code{refuse\_allowed}) & The front end refuses a call that the PDP allowed. \\
Wire & Drop the arguments (\code{drop\_arguments}) & The front end removes the arguments from the request it sends to the PDP. \\
Wire & Drop the destination (\code{drop\_target}) & The front end removes the target from the request it sends to the PDP. \\
Outcome & Flip the decision (\code{flip\_decision}) & The reported decision is inverted. \\
Outcome & Flip one rule verdict (\code{flip\_rule\_verdict}) & One rule verdict in the reported answer is inverted. \\
Outcome & Drop a reason (\code{drop\_reason}) & One reason is removed from the reported answer. \\
Outcome & Swap the argument rewrites (\code{swap\_mutation}) & The reported argument rewrites are replaced by others. \\
Outcome & Flip the effective decision (\code{flip\_effective}) & The reported effective decision is inverted. \\
Outcome & Change the risk level (\code{change\_risk}) & The reported risk level is changed. \\
\bottomrule
\end{tabular}
\end{table}

A case is \emph{applicable} to a fault when the fault can change something observable for it. A fault
that changes nothing for a case is an equivalent fault there, and not a miss. We compute
applicability from the Python engine's verdict on the enforce twin of the case, not from the run:

\begin{itemize}
\item forward a denied call applies when the effective decision is DENY;
\item refuse an allowed call applies when the effective decision is ALLOW or the case has an observe
twin, because an observe twin always allows;
\item drop the arguments applies when the case has arguments;
\item drop the destination applies in egress mode only, because MCP and hook calls carry no target;
\item the six outcome faults apply to every case, because they corrupt every answer they see.
\end{itemize}

For each fault and mode we report the detection rate

\begin{equation}
\text{detection rate} \;=\; \frac{\text{applicable cases flagged by the suite}}{\text{applicable cases evaluated}} .
\label{eq:detection}
\end{equation}

The numerator counts the applicable cases for which at least one axis reported a blocking finding. The
denominator counts the applicable cases that the mode evaluated. We also report which axis flagged each
case. The fault catalogue is the suite's own self-test set. The experiment therefore measures whether
the oracles catch the faults they were built to catch.

\subsection{Front-end source mutants}
\label{sec:m-fesource}

The faults above are injected around the front ends. To change the front ends themselves, we generate
first-order mutants (\cref{tab:operators}) of the functions that build the decision input in the MCP
gateway, in the egress proxy and its body projection, and in the hook bridge's pre-tool path and
tool mapping for the Claude Code hook dialect. Each mutant is installed in memory before the suite
imports the front end. It then runs through the suite's own command line on the packaged set, with
the SDK mode and the mutated mode, against an unmutated PDP. The other two modes are waived by name.
One control run per mode uses no mutant.

We classify each mutant as follows. \emph{Decision differs}: some case's decision, rules or
disposition differ from the SDK mode's. \emph{Evidence only}: only the evidence axis flagged it.
\emph{No decision}: the front end sent a request that the PDP rejected, or produced no decision. The
suite reports this as a mode that could not decide. \emph{Survived}: the suite's verdict stayed
\code{CONFORMANT}, which includes equivalent mutants.

\subsection{Engine mutants}
\label{sec:m-engine}

We generate first-order mutants of the Rego module \code{engine.rego}, of the Python engine and of
the three helpers that the Python engine shares with the policy compiler (address policies, path
normalisation and argument paths). \Cref{tab:operators} lists the mutation operators.

\begin{table}[htbp]
\centering\tablesetup
\caption{Mutation operators and where they were applied.}
\label{tab:operators}
\begin{tabular}{@{}l L{5.3cm} L{4.5cm}@{}}
\toprule
Operator & Change & Applied to \\
\midrule
Relational operator replacement & replace a comparison by another, such as \code{==} by \code{!=} & Rego module; Python engine and helpers; front ends \\
Negation removal & delete a \code{not} & Rego module; Python engine and helpers; front ends \\
And/or swap & replace \code{and} by \code{or} and the reverse & Python engine and helpers; front ends \\
Boolean return flip & a function returns the opposite truth value & Python engine and helpers; front ends \\
Condition deletion & delete one condition of a deny rule & Rego module \\
Field deletion & delete one entry of a dictionary literal or one keyword argument of a constructor call, so that a translator drops a field & front ends \\
\bottomrule
\end{tabular}
\end{table}

Each mutant runs in its own process. A Rego mutant is pushed to a private OPA through its policy
interface. A Python mutant is compiled from the changed syntax tree and installed in memory before
anything imports the module. The suite's runner then evaluates the extended set with its in-process
engines and its reference adapter as the baseline. The front-end modes are waived because this
experiment isolates the engines. The runs use the \NengEvaluated{} extended cases that do not depend on
reasons that only the PDP produces.

For each mutant and case we record four things: whether the two engines' failing-rule sets differ
(the shipped PDP would then answer 503), whether only the argument rewrites differ (the shipped PDP
would still answer), whether an engine raised an error, and whether an expectation failed. These
statements about the shipped PDP are inferred from the engine axis. We confirmed them on a live
stack for one mutant of each of \NliveMutants{} classes (\cref{sec:r-engine}).

\Cref{tab:classes} defines the classes. A mutant is placed in the first class that applies.

\begin{table}[htbp]
\centering\tablesetup
\caption{Outcome classes of an engine mutant, in the order in which they are applied.}
\label{tab:classes}
\begin{tabular}{@{}l L{11.8cm}@{}}
\toprule
Class & Meaning \\
\midrule
Engines disagree & On some case the two engines' failing rules differ. The shipped PDP would answer 503 for requests of that shape. \\
Engine error & An engine raised an error or OPA failed to evaluate. There is no decision. \\
Rewrites differ only & Only the argument rewrites differ. The shipped PDP would answer. \\
Expectation only & The engines agree and a case's expectation fails. \\
Survived & No case distinguishes the mutant. It may be an equivalent mutant. \\
No result & The mutated module cannot be loaded (stillborn), or the test process crashed. These mutants are left out of the rates. \\
Compiler fails & A mutated shared helper breaks the policy compiler, so the policy could not be deployed. \\
\bottomrule
\end{tabular}
\end{table}

We report two rates over the mutants that loaded:

\begin{align}
\text{disagreement rate} &= \frac{\text{mutants with engine disagreement on at least one case}}{\text{mutants that loaded}},
\label{eq:disagreement}\\[1ex]
\text{survival rate} &= \frac{\text{survived mutants}}{\text{mutants that loaded}} .
\label{eq:survival}
\end{align}

Here a mutant that loaded is any mutant that gave a result. Equivalent mutants were not removed, so the
survival rate is an upper bound on the share of undetected defects.

\subsection{Latency}
\label{sec:m-latency}

For latency we use the packaged cases that every mode can express and that need no prior calls
(\NlatCases{} cases, enforce twin). Each run starts a fresh stack and warms each mode up with
\NlatWarmup{} calls. It then makes \NlatRounds{} rounds. In each round the cases are visited in a
shuffled order and, for each case, the four modes in a shuffled order. Every call has its own
principal.

We time the client-side call to the adapter. It includes the front end, which runs in the measuring
process, the PDP reached over the loopback interface, the synchronous append to the audit log in
PostgreSQL, and OPA. We ran \NlatRuns{} runs. For each mode we report the median of the per-run
medians, the range of the per-run medians, and the 90th and 99th percentiles. The extra cost of a
mode over the SDK mode is

\begin{equation}
\Delta_m \;=\; \operatorname{median}_m \;-\; \operatorname{median}_{\text{SDK}} ,
\label{eq:delta}
\end{equation}

where $\operatorname{median}_m$ is the median of the per-run median latencies of mode $m$.

\subsection{Probe of the hook bridge}
\label{sec:m-probe}

To isolate the defect described in \cref{sec:finding}, we also drive the hook bridge alone. A mock PDP
that would deny every call stands in for the real one. We call the bridge with a \code{limit}
argument of $2^{53}-1$, $2^{53}$, $2^{53}+1$, $2^{63}$, $-2^{63}-1$, $10^{30}$, the floating-point number
\code{1e300} and 5, under a fail-closed and a fail-open session. We record how often the PDP was
consulted, what the coding agent was told and what the bridge counted.

\section{Results}
\label{sec:results}

This section answers the four research questions in turn. Each subsection restates the question,
describes the setup, gives the result, explains what the result shows, and ends with a short answer.
The defect found in the hook bridge belongs to the first question and has its own subsection.

\subsection{RQ1: Agreement where both modes can carry the action}
\label{sec:rq1}

\begin{answer}{Question}
When two interception modes can both carry an action, do they produce the same decision, the same
reasons and the same evidence?
\end{answer}

For every case that a native mode could carry, the suite compared the outcome with the SDK mode's
(Property P1, same answer, \cref{eq:equivalence}) and checked the audit chain (Property P2, same
evidence). \Cref{tab:agreement} shows the counts. \emph{Drivable} cases are those that the PDP can
evaluate. \emph{Compared} cases were carried by both the SDK mode and the mode in the row.
\emph{Equal} cases had identical outcomes. \emph{Not expressible} cases could not be carried and are
never counted as equal. \emph{No decision} means that the mode produced no decision at all.

\begin{table}[htbp]
\centering\tablesetup
\caption{Each native mode against the SDK mode (measured). Counts are the same in every run unless a
range is shown.}
\label{tab:agreement}
\begin{tabular}{@{}llrrrrrr@{}}
\toprule
Case set & Mode & Drivable & Compared & Equal & Different & \makecell[r]{Not\\expressible} & \makecell[r]{No\\decision} \\
\midrule
Packaged (5 runs) & MCP & 180 & 87 & 87 & 0 & 93 & 0 \\
 & Egress & 180 & 131 & 131 & 0 & 49 & 0 \\
 & Hook & 180 & 88 & 88 & 0 & 92 & 0 \\
\midrule
Extended (3 runs) & MCP & 1,540 & 345 & 345 & 0 & 1,195 & 0 \\
 & Egress & 1,540 & 1,173 & 1,173 & 0 & 367 & 0 \\
 & Hook & 1,540 & 347 & 347 & 0 & 1,188 & 5 \\
\bottomrule
\end{tabular}
\end{table}

On the packaged set every run was \code{CONFORMANT}: no axis found a blocking difference. On the
extended set, every case that the MCP gateway, the egress proxy or the hook bridge could carry gave the
same decision, the same ordered rule verdicts, the same argument rewrites, the same risk level and the
same controls as the SDK mode. That is \NeqExtPerRun{} comparisons per run, with no difference. The
evidence axis found exactly one agreeing decision event per call in the audit chain, over
\NevidPairsExt{} evidence checks per run.

The exception is \NhookBigintCases{} generated cases on which the hook mode produced no decision at all
(\cref{sec:finding}). They make every extended run \code{DIVERGENT}. The label-only adapters, which
change only the interception-mode field of the request, agreed on all \NwireCompared{} drivable cases,
so the PDP's answer does not depend on that label. The two policy engines never disagreed on any
extended case.

\begin{shows}
Wherever a mode could carry a case, it behaved like the SDK mode, down to the order of the rule
verdicts and the content of the audit record. The one exception is a defect in the hook bridge that
sits before the translation step. Agreement on the cases in the \emph{Equal} column says nothing
about the cases in the \emph{Not expressible} column, which \cref{sec:rq2} examines.
\end{shows}

\begin{answer}{Answer to RQ1}
Yes, on every case that a mode could carry, with one exception. The exception is a defect in the hook
bridge that the generated cases exposed.
\end{answer}

\subsection{A divergence found by the generated cases}
\label{sec:finding}

This subsection describes the exception in detail, because it shows how a front end can fail while
the decision point works correctly.

For \code{limit} set to $2^{53}$, $2^{53}+1$, $2^{63}$, $-2^{63}-1$ or $10^{30}$, the SDK, MCP and egress
modes all decided DENY (rule \code{arguments.max\_value}) and recorded the decision. The hook mode
produced no decision. The bridge refused the call before it consulted the PDP.

\subsubsection{Cause}

The bridge measures the size of the tool input by writing it out with a canonical JSON package that
implements RFC~8785 \cite{rfc8785}. That package accepts only integers in the I-JSON range
\cite{rfc7493}:

\begin{equation}
-(2^{53}-1) \;\le\; n \;\le\; 2^{53}-1 ,
\label{eq:ijson}
\end{equation}

where $n$ is any integer in the tool input. This is the range of integers that a double-precision
number represents exactly. The bridge treats every serialisation error as ``larger than the 4~MiB
limit'', on the stated assumption that the PDP would also refuse such input. The PDP does not. It
accepts these integers and decides. Numbers that are large but representable as doubles, such as
\code{1e300}, are not affected and are decided normally.

\subsubsection{Consequences}

We drove the bridge alone with a mock PDP that would deny every call (\cref{sec:m-probe}).
\Cref{tab:probe} shows the result. The first two rows are the control: an integer inside the I-JSON
range is sent to the PDP, which denies the call and records it. The behaviour for an integer outside the
range depends on the session.

\begin{table}[htbp]
\centering\tablesetup
\caption{The hook bridge driven alone, for an enforce role and a \code{limit} argument (measured, mock
PDP). Zero PDP calls means that no decision and no decision event exist.}
\label{tab:probe}
\begin{tabular}{@{}llrll@{}}
\toprule
\makecell[l]{Integer in\\\texttt{limit}} & \makecell[l]{Session\\fail mode} & \makecell[r]{PDP\\calls} & \makecell[l]{Verdict given to\\the coding agent} & \makecell[l]{Bridge tally} \\
\midrule
$2^{53}-1$ & closed & 1 & deny & 1 decision recorded, 1 call refused \\
$2^{53}-1$ & open & 1 & deny & 1 decision recorded, 1 call refused \\
$2^{53}$ & closed & 0 & deny & 1 call refused \\
$2^{53}$ & open & 0 & none (no verdict) & 1 evidence gap \\
$2^{63}$ & closed & 0 & deny & 1 call refused \\
$2^{63}$ & open & 0 & none (no verdict) & 1 evidence gap \\
$10^{30}$ & closed & 0 & deny & 1 call refused \\
$10^{30}$ & open & 0 & none (no verdict) & 1 evidence gap \\
\bottomrule
\end{tabular}
\end{table}

\begin{itemize}
\item \textbf{Fail-closed session} (an enforce role whose fail mode is \code{closed}, or a session
started with the option \code{-{}-fail-closed}). The call is refused with the reason
\code{argument\_too\_large}, and no decision event is written. The refusal is safe but carries the
wrong reason. A call that the policy would allow, for example with a large integer in an argument that
has no limit, is refused as well.
\item \textbf{Fail-open session} (observe mode, or an enforce role whose effective fail mode is
\code{open\_with\_\allowbreak alert}). The launcher takes the role's fail mode from the PDP, and the
default is \code{open\_with\_\allowbreak alert}. The bridge returns no verdict, writes no decision, tells
the user ``observe mode lets it run'' even in enforce mode, and counts one evidence gap for the session
seal.
\end{itemize}

A hook that stays silent does not approve the call. The call continues through the coding agent's own
permission flow \cite{claudecode-hooks}, and whether it then runs depends on that flow. What the
measurements show is that \sys{} neither stopped the call nor recorded a decision for it.

\begin{shows}
In an enforce session whose role keeps the default fail mode, an agent can move any call onto the
fail-open path by adding one large integer to its arguments. The event is counted but not prevented.
Under the threat model of \cref{sec:threat}, this is a policy bypass (Failure A) whenever the coding
agent's own permission flow then lets the call run, and an unrecorded action or refusal (Failure B) in
every case. The authored cases contained no out-of-range integer, and the system's own probe of
such values compared only the SDK and egress modes. This is why the generated cases found the defect.
\end{shows}

\subsection{RQ2: What each mode can express}
\label{sec:rq2}

\begin{answer}{Question}
When a mode cannot carry what a policy needs, is that reported, and how much of the policy does each
mode actually cover?
\end{answer}

\begin{figure}[htbp]
\centering
\includegraphics[width=\linewidth]{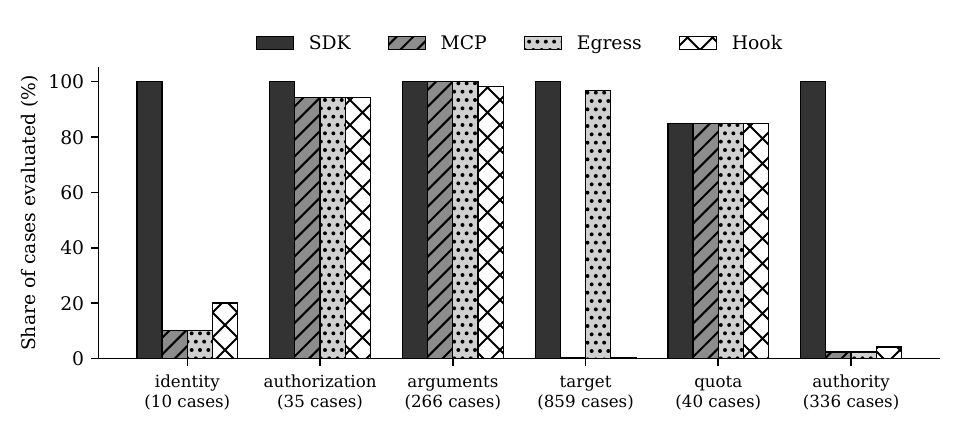}
\caption{Share of extended cases that each mode evaluated, by rule group (measured). The remainder
is reported not expressible, except \NhookBigintCases{} hook cases that produced no decision
(\cref{sec:finding}) and the \NengineOnly{} quota cases that run against the engines only.}
\label{fig:expressible}
\end{figure}

Agreement is only as broad as the set of actions that a mode can carry. For the extended set,
\cref{tab:expressible} and \cref{fig:expressible} show how many of the \NextDrivable{} drivable cases each
mode evaluated, by rule group. \Cref{tab:necounts} lists the reason codes defined in \cref{tab:reasons}
with the number of cases that carry each one.

\begin{table}[htbp]
\centering\tablesetup
\caption{Extended set: cases that each mode evaluated, by rule group (measured). The SDK column leaves
out the \NengineOnly{} cases that run against the engines only.}
\label{tab:expressible}
\begin{tabular}{@{}lrrrrr@{}}
\toprule
Rule group & Cases & SDK & MCP & Egress & Hook \\
\midrule
identity & 10 & 10 & 1 & 1 & 2 \\
authorization & 35 & 35 & 33 & 33 & 33 \\
arguments & 266 & 266 & 266 & 266 & 261 \\
target & 859 & 859 & 3 & 831 & 3 \\
quota & 40 & 34 & 34 & 34 & 34 \\
authority & 336 & 336 & 8 & 8 & 14 \\
\midrule
total & 1,546 & 1,540 & 345 & 1,173 & 347 \\
\bottomrule
\end{tabular}
\end{table}

\begin{table}[htbp]
\centering\tablesetup
\caption{Extended set: number of cases carrying each reason code for \emph{not expressible}, by mode.
A case can carry several codes, and each one is counted.}
\label{tab:necounts}
\begin{tabular}{@{}lrrr@{}}
\toprule
Reason code & MCP & Egress & Hook \\
\midrule
\texttt{auth\_method\_fixed} & 221 & 221 & 109 \\
\texttt{host\_not\_canonical} & -- & 5 & -- \\
\texttt{host\_unconnectable} & -- & 4 & -- \\
\texttt{no\_credential\_expiry} & 277 & 277 & 277 \\
\texttt{no\_delegation\_chain} & 226 & 226 & 226 \\
\texttt{no\_network\_target} & 857 & -- & 857 \\
\texttt{path\_not\_expressible} & -- & 4 & -- \\
\texttt{target\_address\_claim} & -- & 13 & -- \\
\texttt{target\_always\_present} & -- & 13 & -- \\
\texttt{tool\_name\_invalid} & 1 & 1 & 1 \\
\bottomrule
\end{tabular}
\end{table}

Of the \NextDrivable{} drivable cases, the suite reports \NneMcpAll{} not expressible for MCP,
\NneEgressAll{} for egress and \NneHookAll{} for hook. The suite lists each of them with its code, and
none was counted as agreement.

Two different things produce a not-expressible verdict. The first is a limit of the front end as
shipped. Identity is fixed per deployment. No front end carries a credential's expiry time. An HTTP
request carries no delegation chain. The egress proxy refuses non-canonical host spellings and
caller-claimed addresses, because it pins the address itself, which is the property one wants. These
limits account for \NneMcpInherent{} cases for MCP, \NneEgressInherent{} for egress and
\NneHookInherent{} for hook.

The second is a limit of the suite's adapters. The MCP adapter configures no per-tool target rules, so
a tool call carries no network target, and it sends no delegation claim, although the gateway reads a
claimed chain from the \code{\_meta} field of the request. The hook adapter presents every tool as an MCP
tool, and not as the coding agent's own file and fetch tools, from which the bridge derives a path or a
host. A case is adapter-limited when every code on it is one of these limits (\cref{sec:notexpressible}).
This applies to \NneMcpAdapter{} of the \NneMcpAll{} cases for MCP and \NneHookAdapter{} of the
\NneHookAll{} cases for hook. This split comes from reading the source, not from a measurement of those
paths, and a different adapter could change it.

Reporting a limit at test time is not the same as refusing the call at run time. When a call reaches
the PDP without a target, a role's host or address-range (CIDR) restriction is evaluated as \emph{not
applicable} and the call is allowed, also for an allow-list. Two packaged cases assert exactly this, and
the SDK, MCP and hook modes agree on it. The decision record shows that the rule did not apply, so the
gap is visible in the evidence. An operator who deploys a destination allow-list behind an MCP gateway
that has no target rules nevertheless gets no destination control there.

\begin{shows}
The hook mode could not carry \NneHookPctExt\% of the drivable cases, and the MCP mode could not
carry \NneMcpAll{} of \NextDrivable{}. Most of the MCP and hook gap comes from how the suite drives
them, not from the gateway or the bridge. The stronger statement is therefore that the suite does not
yet exercise the target and delegation rules through the MCP and hook front ends. The egress mode
covers the most (\NneEgressAll{} not expressible), and all of its gaps are limits of the proxy.
\end{shows}

\begin{answer}{Answer to RQ2}
The suite reports every case that a mode cannot carry, with a reason, and never counts it as equal.
Outside the SDK mode, a large part of the policy is not exercised, most of it in the target and
authority rule groups. Much of the gap is due to the adapters, which the suite can close.
\end{answer}

\subsection{RQ3: What the tests catch}
\label{sec:rq3}

\begin{answer}{Question}
How much can the tests detect: comparing modes, comparing the two policy engines, and checking the
expected result of each case? What do they miss by construction?
\end{answer}

We answer in three steps: faults injected around the front ends, mutants of the front ends' own
source, and mutants of the policy engines.

\subsubsection{Injected front-end faults}
\label{sec:r-faults}

\Cref{tab:faults} gives, for each fault of \cref{tab:faultdefs} and each mode, the number of applicable
cases that the suite flagged, out of all applicable cases, and the axes that flagged them
(\cref{sec:m-faults}). The control run, with no fault, flagged no case.

\begin{table}[htbp]
\centering\tablesetup
\caption{Injected faults, applied to every call of one mode over the packaged cases (measured). Each
cell shows cases flagged / applicable cases. n/a: the fault cannot change any case in that mode,
because an MCP or hook call carries no target.}
\label{tab:faults}
\begin{tabular}{@{}lrrrL{5.2cm}@{}}
\toprule
Injected fault & \multicolumn{3}{c}{Flagged / applicable} & Flagged by axes \\
\cmidrule(lr){2-4}
 & MCP & Egress & Hook & \\
\midrule
\multicolumn{5}{@{}l}{\emph{Faults on the exchange with the decision point}} \\
\hspace{1em}Forward a denied call & 45/45 & 79/79 & 46/46 & cross-mode, evidence, observe/enforce \\
\hspace{1em}Refuse an allowed call & 87/87 & 131/131 & 88/88 & cross-mode, evidence, observe/enforce \\
\hspace{1em}Drop the arguments & 72/72 & 72/72 & 73/73 & cross-mode, evidence \\
\hspace{1em}Drop the destination & n/a & 131/131 & n/a & cross-mode, evidence \\
\midrule
\multicolumn{5}{@{}l}{\emph{Faults in the outcome the front end reports}} \\
\hspace{1em}Flip the decision & 87/87 & 131/131 & 88/88 & cross-mode, observe/enforce \\
\hspace{1em}Flip one rule verdict & 87/87 & 131/131 & 88/88 & cross-mode \\
\hspace{1em}Drop a reason & 87/87 & 131/131 & 88/88 & cross-mode \\
\hspace{1em}Swap the argument rewrites & 87/87 & 131/131 & 88/88 & cross-mode \\
\hspace{1em}Flip the effective decision & 87/87 & 131/131 & 88/88 & cross-mode, observe/enforce \\
\hspace{1em}Change the risk level & 87/87 & 131/131 & 88/88 & cross-mode \\
\midrule
Control (no fault) & \multicolumn{3}{c}{0 cases flagged} & \\
\bottomrule
\end{tabular}
\end{table}

The suite flagged \NfeCaught{} of \NfeApplicable{} applicable pairs of case and fault. Applicability
matters. Dropping the destination on the MCP gateway changes nothing, because an MCP call carries no
target. Forwarding a denied call changes nothing for the two cases with an unknown or case-mismatched
role, which the policy runs in its default observe mode, so the effective decision is already ALLOW.
These are equivalent faults for those cases and not misses.

The axis that catches a fault differs by fault. A dropped argument made the cross-mode comparison fail
on only \NdropArgsX{} of \NdropArgs{} applicable cases. The other \NdropArgsEvOnly{} were caught only by
the evidence axis, which compares the request that the front end sent to the PDP with what the caller
supplied. A suite that compared answers alone would have passed a translator that loses arguments
whenever the lost argument does not matter to the current policy, and would have failed the day it
does. The same holds for a proxy that loses the destination: all \NdtCases{} cases were flagged by the
evidence axis, and \NdtCross{} by the cross-mode comparison.

For refusing an allowed call, the evidence axis flagged every case in every mode. The observe/enforce
axis flagged all but \NrefOeMissed{} cases per mode. The cross-mode comparison flagged fewer than half
of the cases, because it sees only the enforce twins. The six outcome faults are flagged on every case by
the cross-mode comparison, as they must be, because they corrupt a field that the comparison reads. They
show that the comparison is not blind to any reported field, not that it is hard to fool.

\begin{shows}
Every injected fault that could change something was flagged. Which axis noticed it differs, and
several faults would have been missed by a suite that compared only the decisions. This is why the
suite records and compares the request that each front end sent to the PDP. The faults come from the
suite's own self-test catalogue, so the result shows that the tests catch the faults they were built to
catch, not that they catch every fault.
\end{shows}

\subsubsection{Front-end source mutants}
\label{sec:r-fesource}

\Cref{tab:fesource} summarises the \NfeMutants{} mutants of the front ends' source
(\cref{sec:m-fesource}).

\begin{table}[htbp]
\centering\tablesetup
\caption{Source mutants of the front ends' request builders and of the hook bridge's pre-tool path,
run through the live suite on the packaged cases (measured). \emph{Decision differs}: some case's
decision, rules or disposition differ from the SDK mode's. \emph{Evidence only}: only the evidence
axis flagged it. \emph{Survived}: the verdict stayed \code{CONFORMANT}, including equivalent mutants.}
\label{tab:fesource}
\begin{tabular}{@{}lrrrrr@{}}
\toprule
Mutated code & \makecell[r]{Decision\\differs} & \makecell[r]{Evidence\\only} & \makecell[r]{No decision\\(error)} & \makecell[r]{Survived} & Total \\
\midrule
MCP gateway: request builder & 1 & 0 & 14 & 14 & 29 \\
Egress proxy: request builder & 5 & 1 & 15 & 10 & 31 \\
Egress proxy: body projection & 3 & 0 & 4 & 13 & 20 \\
Hook bridge: pre-tool path & 3 & 7 & 19 & 23 & 52 \\
Hook bridge: tool mapping & 1 & 0 & 3 & 13 & 17 \\
\midrule
All front-end mutants & 13 & 8 & 55 & 73 & 149 \\
\bottomrule
\end{tabular}
\end{table}

Most of the mutants that produced no decision delete a required field such as the role, the identity,
the tool or the context. The PDP rejects such a request, and the suite reports a mode that could not
decide. The mutants that only the evidence axis caught include a bridge that drops the observe-mode
warning shown to the user, and one that returns the original arguments instead of the rewritten ones.

We classified the survivors by reading the code at each site, which is a judgement and not a proof
(Appendix~\ref{app:survivors}). Four groups are explained by redundancy or by what the adapters and cases can
reach: code that the adapters never drive, fields that no rule reads, parser hardening that
dictionary-level cases cannot exercise, and fields that are carried twice. The fifth group contains two
real gaps in what the packaged cases reach.

\begin{itemize}
\item Deleting \code{authority\_source} from the request that the MCP gateway or the egress proxy
builds survives. Every packaged case that sets an authority source also delegates, and delegation is
not expressible in those modes. A front end could drop the field that records on whose authority a
call is made, and the packaged suite would not notice. Generated authority cases that a front end can
carry, namely an authority source without delegation, do flag both deletions.
\item Deleting \code{acting\_for} or \code{delegation\_depth} from the MCP gateway's request survives,
although the gateway reads a claimed chain from the \code{\_meta} field. The suite never sends one, so
these deletions remain unreached.
\end{itemize}

\begin{shows}
The suite noticed about half of the front-end mutants: \NfsErr{} because the PDP rejected the request,
\NfsDec{} because a decision, rule or disposition changed, and \NfsEv{} only through the evidence axis. The
survivors map the parts of the front ends that the packaged cases do not reach. Two matter: the
authority source and the delegation fields. The evidence axis was needed to notice the \NfsEv{} mutants
that no decision comparison would have flagged.
\end{shows}

\subsubsection{Engine mutants}
\label{sec:r-engine}

\begin{figure}[htbp]
\centering
\includegraphics[width=\linewidth]{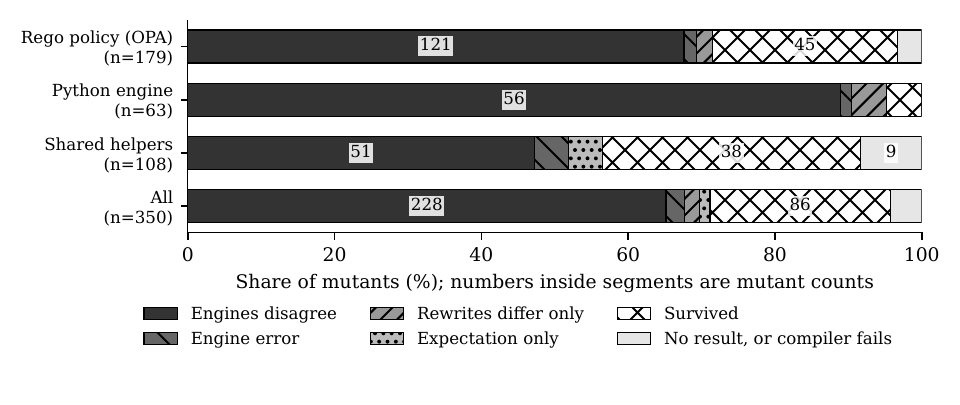}
\caption{Outcome of the engine mutants as a share of each target group (measured). Numbers inside
the segments are mutant counts.}
\label{fig:mutants}
\end{figure}

\Cref{tab:mutants} and \cref{fig:mutants} classify the \NengMutants{} engine mutants by the strongest
signal that they produced (\cref{tab:classes}).

\begin{table}[htbp]
\centering\tablesetup
\caption{First-order engine mutants, by the strongest signal over the extended cases (measured). The
meaning of each class is in \cref{tab:classes}. ``Engines disagree'' means that the shipped PDP would
answer 503.}
\label{tab:mutants}
{\setlength{\tabcolsep}{3.2pt}\footnotesize\begin{tabular}{@{}lrrrrrrrr@{}}
\toprule
Mutated code & \makecell[r]{Engines\\disagree\\(503)} & \makecell[r]{Engine\\error} & \makecell[r]{Rewrites\\differ\\only} & \makecell[r]{Expect-\\ation\\only} & \makecell[r]{Survived} & \makecell[r]{Compiler\\fails} & \makecell[r]{No\\result} & Total \\
\midrule
Rego policy (OPA) & 121 & 3 & 4 & 0 & 45 & 0 & 6 & 179 \\
Python engine & 56 & 1 & 3 & 0 & 3 & 0 & 0 & 63 \\
Shared helpers & 51 & 5 & 0 & 5 & 38 & 8 & 1 & 108 \\
\midrule
All engine mutants & 228 & 9 & 7 & 5 & 86 & 8 & 7 & 350 \\
\bottomrule
\end{tabular}}
\end{table}

Of the \NmLive{} mutants that loaded, \NmAllSplit{} (\NmSplitPct\%) made the two engines' failing rules
differ on at least one case. The shipped PDP would therefore have answered 503 for requests of that
shape instead of deciding. The median such mutant splits \NmSplitMedianCases{} of the
\NengEvaluated{} evaluated cases, and \NmSplitOneCase{} split exactly one. A further \NmAllErr{} made an
engine raise an error or OPA fail to evaluate, which again means no decision. The policy compiler failed
for \NmAllComp{} mutants of shared helpers, so the policy could not have been deployed. These outcomes
fail closed: the defect stops some decisions instead of changing them.

The other outcomes are weaker. In \NmAllMutOnly{} mutants only an argument rewrite changed. The shipped
check would have answered, and only the suite's engine axis, which also compares argument rewrites,
flagged them. In \NmAllExpOnly{} mutants, all in shared helpers, the engines still agreed, and only a
case expectation caught them. Finally, \NmAllSurv{} mutants (\NmSurvPct\%) survived every case.

Most surviving Rego mutants are equivalent, or they would matter only for a combination of inputs or a
boundary value that no case contains. Shared-helper survivors (\NmSharedSurv{} of \NmSharedTotal{}) are in
the argument-path helpers, the address parser and the path normaliser, where both engines consume the
helper's output and no expectation pins the behaviour (Appendix~\ref{app:survivors}).

We confirmed this inference on a live stack for one mutant of each of \NliveMutants{} classes. A Rego
mutant that splits every request made the PDP answer 503 to every call in every mode, and the suite reported each as a mode that could not
decide. A Rego mutant that changes only argument rewrites passed the cross-mode, expectation and
evidence axes in all four modes, because the PDP applies the Python engine's rewrites, and only the
engine axis flagged it.

\subsubsection{Two blind spots of two-engine consensus}
\label{sec:blind}

\begin{itemize}
\item \textbf{A common-mode fault is invisible to consensus.} A defect in a helper that both engines
rely on changes both answers in the same way, so consensus cannot see it. Only an expectation can. The
data bound this blind spot but do not prove it. Of \NmSharedTotal{} shared-helper mutants,
\NmSharedSplit{} still made the engines disagree, \NmSharedExpOnly{} were seen only by an expectation,
and \NmSharedSurv{} survived everything. An unknown number of these survivors are equivalent.
\item \textbf{The shipped check ignores argument rewrites.} The PDP compares the two engines' failing-rule
sets, not their argument rewrites. A fault that changes only an argument rewrite, for example a field
that should be removed and is not, passes the shipped check and returns a decision. The suite's engine
axis, which compares rewrites as well, flags it.
\end{itemize}

Both blind spots follow from design choices that make the engines agree. Both are visible only with
expectations or a stricter comparison.

\subsubsection{Generated cases against engine mutants}

The packaged cases alone flagged \NmFlagPkg{} engine mutants. Adding the \NgenCases{} generated cases
raised this to \NmFlagExt{}. The \NmFlagGenOnly{} additional mutants are in the equality function for JSON
values (type distinctions such as \code{0} against \code{false}), the path decoder and the
argument-path walker, which the generated value shapes reach. Consistency-only cases cannot catch a
common-mode fault by construction, and a fault confined to one engine is usually caught by the authored
set already. The generated cases were more useful against the front ends: they found the defect of
\cref{sec:finding} and they reach the \code{authority\_source} mutants above.

\begin{shows}
Two-engine consensus exposes most faults that are confined to one engine. For \NmSplitPct\% of loaded
mutants the engines disagreed on at least one case, and the decision point would then refuse to decide
for requests of that shape. Consensus does not help when both engines share the defect, and the shipped
check does not compare argument rewrites. Expectations that a person wrote remain necessary. The
generated consistency-only cases, which have no expectations, do not replace them.
\end{shows}

\begin{answer}{Answer to RQ3}
The tests flag every injected fault that could change something. They leave \NmSurvPct\% of the loaded
engine mutants and \NfsSurv{} of the \NfeMutants{} front-end mutants unnoticed, and some of these are
equivalent mutants. Comparing decisions alone is not enough: the evidence comparison catches
translation faults that leave the decision unchanged. Two-engine consensus has two structural blind
spots, defects in shared code and changes to argument rewrites alone, which only a written
expectation or a stricter comparison can cover. The results show detection of the faults that we
thought of, not detection of all faults.
\end{answer}

\subsection{RQ4: Latency of each mode}
\label{sec:rq4}

\begin{answer}{Question}
How much decision latency does each interception mode add?
\end{answer}

\begin{figure}[htbp]
\centering
\includegraphics[width=\linewidth]{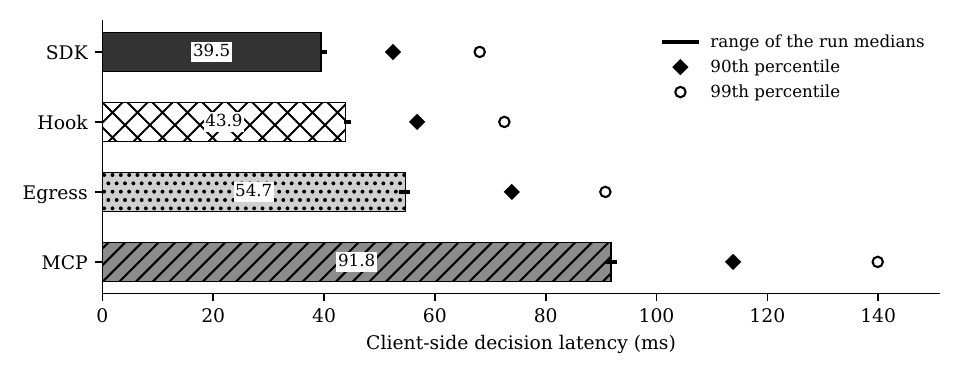}
\caption{Median latency per mode, with the range of the per-run medians and the 90th and 99th
percentiles (measured).}
\label{fig:latency}
\end{figure}

\Cref{tab:latency} and \cref{fig:latency} show the client-side latency of each mode over \NlatRuns{} runs
of \NlatCallsPerRun{} interleaved calls (\cref{sec:m-latency}). All times are in milliseconds. They
include the front end, the PDP, the audit append and OPA.

\begin{table}[htbp]
\centering\tablesetup
\caption{Client-side decision latency per mode in milliseconds (measured). The median is the median of
the per-run medians; the next column gives the range of those medians. The last two columns give the
median for allowed and for denied calls.}
\label{tab:latency}
\begin{tabular}{@{}lrrrrrrr@{}}
\toprule
Mode & Median & \makecell[r]{Range of\\run medians} & \makecell[r]{90th\\percentile} & \makecell[r]{99th\\percentile} & \makecell[r]{Extra over\\SDK} & \makecell[r]{Allowed\\calls} & \makecell[r]{Denied\\calls} \\
\midrule
SDK & 39.5 & 39.3--40.6 & 52.5 & 68.1 & -- & 39.2 & 39.8 \\
MCP & 91.8 & 90.8--92.9 & 113.8 & 139.9 & +52.3 & 95.3 & 89.3 \\
Egress & 54.7 & 53.5--55.6 & 73.9 & 90.8 & +15.2 & 56.7 & 52.7 \\
Hook & 43.9 & 43.6--44.9 & 56.8 & 72.5 & +4.4 & 43.6 & 44.3 \\
\bottomrule
\end{tabular}
\end{table}

The per-run median varies by at most \NlatSpreadMax{}~ms within a mode, so the differences between modes
are larger than the differences between runs. The SDK mode makes one loopback HTTP call to the PDP, which
includes the synchronous, chained audit append to PostgreSQL, and took a median of \NlatPSdk{}~ms. The
hook bridge added little (\NlatPHook{}~ms): it builds the request in process and makes the same single
call. The egress proxy (\NlatPEgress{}~ms) and especially the MCP gateway (\NlatPMcp{}~ms) add a network hop
and their own protocol handling.

An allowed call through MCP or egress also includes forwarding to the stub tool or target. The
denied path forwards nothing, and shows nearly the same cost (\NlatDMcp{}~ms and \NlatDEgress{}~ms,
against \NlatDSdk{}~ms for the SDK mode). Most of the cost therefore belongs to the front end itself.
In the same runs the four modes' decisions disagreed in \NlatDisagree{} of the \NlatGroups{} groups of
the same case and round.

\begin{shows}
Choosing a mode trades coverage against cost. The hook adds a few milliseconds to the SDK path, the
egress proxy adds more, and the MCP gateway adds the most, more than twice the SDK path's median. The
front ends run in the same process as the timing client, which probably inflates their figures. The
numbers show the order of the per-mode cost on one development build and one laptop, not what a tuned
deployment would achieve. The hook figure is the cost of the bridge alone and excludes the coding
agent's own cost of invoking the hook.
\end{shows}

\begin{answer}{Answer to RQ4}
On this machine the median decision took between \NlatPSdk{}~ms (SDK) and \NlatPMcp{}~ms (MCP). The hook
bridge adds the least to the SDK path and the MCP gateway the most.
\end{answer}

\section{Discussion}
\label{sec:discussion}

This section interprets the results and draws practical conclusions. The statements are limited to
what the measurements show for one system (see \cref{sec:limitations}).

\subsection{Agreement is cheap where it is defined}

All four modes reach one decision point, and that decision point does not read the mode label, as the
label-only adapters confirmed. Agreement on what both sides can express is then mostly a property of the
front ends' translators. The suite's witnesses (wire tap, protected upstream and client view) check
these directly.

The one divergence we found was not in a translator's mapping. It was in a guard that the hook bridge
applies \emph{before} translation, with its own idea of what the PDP accepts. In this system, such
pre-checks were where differential testing paid off. They are a sensible place to look first in other
systems.

\subsection{``Not expressible'' is partly a statement about the tests}

The MCP and hook modes carried well under half of the extended cases. Some of the gap is structural:
the shipped front ends fix identity per deployment, and an HTTP request carries no delegation chain.
A larger part, as the suite drives them, is adapter configuration. This concerns target rules, the
coding agent's own tools and a delegation claim in the MCP request.

An operator who writes a target or authority rule and deploys only a gateway should verify that the
rule can fire there. The suite reports this for each case. A deployment-time check that maps each rule
to the modes that can ever trigger it would make the same point earlier.

\subsection{Comparing evidence finds faults that decisions hide}

For \NdropArgsEvOnly{} of \NdropArgs{} dropped-argument cases, the decision did not change, and only the
comparison of the recorded request noticed the loss. A front end that drops an argument is harmless
until a policy depends on that argument, and then it fails without any sign in the decision. Comparing
the request that each front end sent exposes such a fault early.

\subsection{Two engines are not two specifications}

Consensus caught faults that were confined to one engine. It cannot catch a fault in the parts that both
engines share by construction, and the shipped check deliberately ignores argument rewrites.
Expectations written by a person remain necessary. The generated cases, which carry no expectations, do
not replace them.

\subsection{The fail mode decides how serious the defect is}

The hook defect is a refusal with the wrong reason in a fail-closed session, and a missing decision in
a fail-open one. Enforce-mode roles start with the fail-open default. Setting the fail mode to
\code{closed} is therefore a precondition for treating enforce mode as a control.

\subsection{Practical guidance}

\subsubsection{For operators}

\begin{itemize}
\item Set the fail mode of every enforce-mode role to \code{closed}.
\item Check, for each interception point that you deploy, that every rule you rely on can fire there. A
destination allow-list behind an MCP gateway without target rules gives no destination control.
\item Read the evidence-gap counter of a session. It marks calls that were not decided.
\end{itemize}

\subsubsection{For builders of governance layers}

\begin{itemize}
\item Run every shipped front end against one baseline. Put a recording transport on the exchange
with the decision point and a protected upstream that records whether the tool ran.
\item Compare the request that each front end sent, and not only the decision.
\item Review the checks that front ends apply before they ask the decision point, such as size limits
and parsers. They carry their own assumptions about what the decision point accepts.
\item Report what a front end cannot express as a result in its own right, with a reason.
\end{itemize}

\subsubsection{For testers}

\begin{itemize}
\item Keep expectations written by a person. Consistency-only cases cannot catch common-mode faults.
\item Mutate both the engines and the code they share, and compare argument rewrites as well as
failing rules.
\end{itemize}

\section{Limitations and future work}
\label{sec:limitations}

\subsection{Limitations}

\begin{itemize}
\item \textbf{One system, built and tested by its author.} The system, its test suite and adapters, the
generated cases, the fault and mutant catalogues and this analysis were all produced by one person. The
oracle-power results therefore show detection of the faults we thought of, and no other governance layer
was tested. \emph{How to address it:} replicate the measurement on independently built governance
layers, with fault catalogues written by other people.
\item \textbf{Synthetic inputs and simplified drivers.} The cases are synthetic requests against a
purpose-built policy. The generated grid is systematic but not exhaustive, and it uses IP literals, so
resolution through the Domain Name System, rebinding and name canonicalisation at the proxy were not
exercised. The SDK baseline is the wire call and not the SDK's framework wrappers. The MCP and hook
adapters configure no target rules and send no delegation claim, and no real coding agent is involved.
\emph{How to address it:} extend the adapters to drive these paths, add DNS-based cases, and draw cases
from recorded agent traces.
\item \textbf{First-order mutants and inferred outcomes.} Each mutant contains one change from a small
operator set, and equivalent mutants were not removed, so survival is an upper bound on missed faults.
Survivors were classified by reading, and the statement that the shipped PDP would answer 503 was
confirmed on a live stack for \NliveMutants{} mutants and inferred for the rest. \emph{How to address
it:} use higher-order mutants, detect equivalent mutants automatically, and run every mutant live.
\item \textbf{The hook defect was shown on the bridge alone.} Its fail-open consequence was measured
with a mock PDP, and whether a coding agent then runs the call depends on its permission flow.
\emph{How to address it:} test the hook bridge with a real coding agent.
\item \textbf{One machine and one build.} Latency comes from one laptop over the loopback interface,
with a synchronous audit append and the front ends in the measuring process. It is not a capacity
figure. The database and cache images are referenced by tag, not by digest. \emph{How to address it:}
repeat the measurements on dedicated hardware, with separate processes and images pinned by digest.
\end{itemize}

\subsection{Future work}

\begin{itemize}
\item Make the shipped consensus check compare argument rewrites, as the suite's engine axis does.
\item Add cases that combine an unknown role with each identity and authority violation, and cases at
the exact limits on path length and segment count.
\item Let the MCP and hook adapters drive per-tool target rules, the delegation claim and the coding
agent's own file and fetch tools, so that target and authority rules are exercised through those modes.
\item Add a deployment-time check that lists, for each rule of a policy, the modes that can ever trigger it.
\end{itemize}

\section{Data and artifact availability}
\label{sec:availability}

\subsection{What is released}

The harness, the cases, the raw results and the scripts that regenerate every table and number in this
paper are distributed as ancillary files of this arXiv submission. They are in the folder \code{anc/} of
the upload, and the abstract page of the paper lists them as ``Ancillary files''. The ancillary files
are one archive, \code{p6-artifact.tar.gz}. It contains:

\begin{itemize}
\item the harness: the four-mode runner, the timing wrapper around the suite's adapters, the fault and
mutation drivers, the latency driver and the probe of the hook bridge;
\item the generator of the extended case set and the generated cases;
\item the analysis script \code{harness/analyze.py}, which produces every table, figure and number of
this paper from the raw data;
\item the independent re-check script \code{harness/independent\_check.py}, which imports neither the
analysis script nor the system and recomputes every figure from the raw results;
\item the raw results: suite reports, timing, mutant records, logs and environment files.
\end{itemize}

Local paths and user names were removed from the raw results before release, and nothing else was
changed.%
\zenododoi{TO-BE-ADDED}

\subsection{How to recompute the numbers}

The independent re-check needs only Python 3.11 or later and PyYAML. From the folder that holds the
unpacked archive, the command is

\begin{Verbatim}[fontsize=\small,xleftmargin=2em]
python -I harness/independent_check.py
\end{Verbatim}

It re-derives every number that the paper quotes from the raw files, compares each one with the value in
the paper, and prints one line per check. It finished in about five seconds on the machine of
\cref{tab:environment}. The analysis script regenerates the tables, the figures and the file of numbers
with \code{python harness/analyze.py}. It loads the case sets with the system's own loader, so it needs
the system's source code.

\subsection{The system's source code}

The source code of \sys{} will be released under the Apache-2.0 licence with the product's first public
release. Until then, the experiments cannot be repeated without it. The archive nevertheless contains
everything needed to recompute the tables and numbers from the raw data, and to inspect the cases, the
harness and the raw results. Appendix~\ref{app:repro} lists the commands that repeat the experiments once the
source is available, and the time each takes.

\section{Conclusion}
\label{sec:conclusion}

We tested whether one policy means the same thing at the four interception points of one governance
system. On every case that a front end could carry, the decision, the reasons and the audit record
equalled the SDK path's. The exception was a defect in a pre-check of the hook bridge, which a grid of
generated cases exposed. It let an agent-chosen argument move a call onto the fail-open path.

As the suite drives them, a large part of the policy is not exercised outside the SDK path, and much of
that is due to the adapters. Reporting this for each case, with a reason, is the honest form of an
equivalence claim. Two-engine consensus catches faults confined to one engine, but it is blind to some
faults in shared code and, in the shipped check, to changes of argument rewrites alone. Written
expectations remain necessary.

The method transfers to other systems: run every shipped front end against one baseline, account for
what each cannot express and why, compare the recorded request and not only the decision, and mutate
both the engines and the code they share.

\section*{Acknowledgements}
The author designed the study, ran and checked the experiments, and is responsible for all content.
Generative AI assistants were used for drafting and coding support.

\bibliographystyle{IEEEtran}
\bibliography{refs}

\appendix
\section{Glossary of terms and abbreviations}
\label{app:glossary}

\small
\setlength{\LTpre}{4pt}
\begin{longtable}{@{}L{3.9cm} L{12.0cm}@{}}
\toprule
Term & Meaning \\
\midrule
\endfirsthead
\toprule
Term & Meaning \\
\midrule
\endhead
\bottomrule
\endlastfoot
Adapter & Test code that presents a case to the PDP through one interception mode and returns the outcome. \\
Adapter-limited & A case that a mode cannot carry only because of how the suite's adapter is configured (\cref{sec:notexpressible}). \\
Agent, AI agent & A program in which a large language model chooses what to do next and acts through tools. \\
ALLOW, DENY & The two decisions of the PDP. \\
Argument & A value passed to a tool, such as a search text or a limit. \\
Argument rewrite & A change that the PDP tells the caller to apply to the arguments, such as removing an access key. The system's data model calls it a mutation. \\
Audit event & The record of one decision, sealed by the PDP before it answers. \\
Axis & One family of checks of the conformance suite (\cref{tab:axes}). \\
Baseline & The SDK mode, against which the other modes are compared. \\
Case & A recorded decision input with an optional expectation. The suite calls it a fixture. \\
CIDR & Classless Inter-Domain Routing notation for a range of network addresses, such as \code{10.0.0.0/8}. \\
Common-mode fault & A defect in code that both policy engines share, so that both give the same wrong answer. \\
Conformance suite & The test package that ships with the system. It replays cases through all modes and compares the results. \\
Control run & A run with no fault or mutant, used to check that the harness itself reports nothing. \\
Decision input & The description of a proposed tool call that is sent to the PDP. \\
Differential testing & Giving the same input to systems that should behave alike and comparing their outputs. \\
DNS & Domain Name System, which maps host names to network addresses. \\
Effective decision & What actually happens to the action. It equals the decision, except in observe mode, where it is always ALLOW. \\
Egress proxy & An HTTP proxy on the agent's network path that asks the PDP about each request. \\
Engine consensus & The check that the Python engine and OPA agree. The PDP answers 503 when they do not. \\
Enforce mode & A role mode in which an action proceeds only if the decision is ALLOW. \\
Equivalent mutant & A mutant whose change does not alter behaviour, so no test can distinguish it. \\
Evidence gap & A call for which no decision record exists. \\
Expectation & The result that a case's author intended: the decision and the failing rules. \\
Expressible & A case that a mode can present to the PDP without changing what it asks. \\
Extended set & The packaged set plus the generated cases. \\
Fail mode & What happens when no usable decision is available: \code{closed} refuses the call, \code{open\_with\_alert} lets it proceed and raises an alert. \\
Fail-closed, fail-open & Behaviour in which a call is refused, or allowed to proceed, when no decision is available. \\
Fault injection & Breaking a component on purpose to see whether the tests notice. \\
First-order mutant & A mutant that contains exactly one change. \\
Front end & The code at an interception point that talks to the PDP. \\
Governance layer & Software that checks each tool call against a policy before it runs and records the decision. \\
Hash chain & A log in which each event contains the hash of the previous one, so that changes to earlier events can be detected. \\
Hook, hook bridge & A program that a coding agent runs before each tool call. The bridge answers it by asking the PDP. \\
HTTP & Hypertext Transfer Protocol. \\
I-JSON & A restricted form of JSON for interoperability. It limits integers to the range $-(2^{53}-1)$ to $2^{53}-1$ \cite{rfc7493}. \\
IPv4, IPv6 & The two versions of the Internet Protocol, which give network addresses such as \code{192.0.2.1} and \code{2001:db8::1}. \\
Interception mode, point & A place where an action can be checked, and the configuration that uses it: SDK, MCP, egress or hook. \\
JSON & JavaScript Object Notation, a text format for structured data. \\
MCP & Model Context Protocol, a protocol that connects agents to tool servers \cite{mcp-spec-2025-06-18}. \\
MCP gateway & A gateway between an MCP client and an MCP server that asks the PDP about each tool call. \\
Mutant & A copy of source code with one small change. It is killed if a test notices it and survives if none does. \\
NAT64, 6to4 & IPv6 address forms that embed an IPv4 address. \\
Native mode & One of the three modes whose front end is a real component: MCP, egress and hook. \\
Not expressible & The state of a case that a mode cannot carry. It is never counted as equal. \\
Observe mode & A role mode in which the decision is recorded but the action always proceeds. \\
OIDC & OpenID Connect, an identity protocol. \\
OPA & Open Policy Agent, a policy engine \cite{opa-rest-api}. \\
Oracle & The rule that decides whether an observed result is correct. \\
Outcome & The eight fields of a decision that the suite compares (\cref{sec:property}). \\
Packaged set & The cases that ship with the conformance suite. \\
PDP & Policy decision point, the service that evaluates the policy for each decision input. \\
Policy & The operator's rules, written per role. \\
Prompt injection & Text that the agent reads and that contains instructions steering the model. \\
Quota & A rule group that limits the rate of calls and detects loops. \\
Reason code & A stable name for why a mode cannot carry a case (\cref{tab:reasons}). \\
Reference engine & The Python implementation of the policy, evaluated next to OPA. \\
Rego & The policy language of OPA. \\
RFC & Request for Comments, a document of the Internet standards process. \\
Role & A named set of permissions to which the policy attaches rules. \\
SDK & Software development kit. Here it is a library that wraps a tool function and asks the PDP. \\
Stillborn mutant & A mutant that cannot be loaded. \\
Survived (mutant) & A mutant that no test notices. \\
Target & The destination of a call: a network host or address, or a file path. \\
Tool, tool call & A function or service that an agent uses, and one use of it with its arguments. \\
Twin roles & The enforce and observe copies of a role, identical except for their mode. \\
Verdict & The result of one rule (pass, fail or not applicable), or the overall result of a suite run (\code{CONFORMANT}, \code{DIVERGENT}, \code{INCOMPLETE} or \code{ERROR}). \\
Wire tap & A recording transport on the exchange between a front end and the PDP. \\
YAML & A text format for structured data, used for policies and cases. \\
\end{longtable}
\addtocounter{table}{-1}
\normalsize

\section{Re-running the experiments}
\label{app:repro}

Re-running the experiments needs the system's source code (\cref{sec:availability}), Python 3.13,
Docker and the images named in \cref{tab:environment}. Set the environment variable
\code{OATHPOST\_REPO} to the location of the source. The commands below are run from the folder of the
unpacked archive. The harness creates and removes its own containers and never touches another stack.

\begin{Verbatim}[fontsize=\footnotesize,xleftmargin=1em]
python harness/infra.py up
python fixtures/generate.py
python harness/run_conformance.py --label packaged --repeat 5
python harness/run_conformance.py --label extended --repeat 3 \
    --fixtures fixtures/extended
python harness/frontend_faults.py
python harness/engine_mutation.py --fixtures fixtures/extended --label engines
python harness/live_mutants.py
python harness/frontend_mutation.py
python harness/frontend_mutation.py --label frontend-source-followup \
    --fixtures fixtures/extended --extra --group authority \
    --only control-mcp control-egress "mcp_gateway.gateway#184" "egress.proxy#249"
python harness/probe_hook_bigint.py
python harness/probe_unknown_role_fault.py
python harness/latency.py --runs 5 --rounds 10
python harness/analyze.py
python harness/independent_check.py
python harness/infra.py down
\end{Verbatim}

\Cref{tab:durations} gives the time that each step took on the machine of \cref{tab:environment}.

\begin{table}[htbp]
\centering\tablesetup
\caption{Time taken by each step on the test machine (measured). The two mutant campaigns used two
parallel workers, so their elapsed time was shorter than the worker time shown.}
\label{tab:durations}
\begin{tabular}{@{}l l@{}}
\toprule
Step & Time \\
\midrule
One packaged run & one to two minutes \\
One extended run & about \NextWallMin{} minutes \\
Injected-fault campaign (\NfeRunsAll{} runs) & \NfaultHours{} worker-hours \\
Front-end source mutants (\NfeMutants{}) & \NfeSrcHours{} worker-hours \\
Engine mutants (\NengMutants{}) & \NengHours{} worker-hours \\
Analysis script & under one minute \\
Independent re-check & about five seconds \\
\bottomrule
\end{tabular}
\end{table}

\section{Surviving mutants}
\label{app:survivors}

This appendix gives the author's classification of the mutants that survived every case. The
classification comes from reading the code at each site. It is a judgement and not a proof.

\subsection{Front-end mutants}

The \NfsSurv{} surviving front-end mutants fall into five groups.

\begin{enumerate}[label=(\roman*)]
\item \textbf{Code that the adapters never drive.} The hook adapter presents every tool as an MCP tool.
Mutants in the mapping of the agent's own shell, file and web tools, in the gateway-routed hook path and
in response branches that need rewriting or auto-allow configuration cannot be reached.
\item \textbf{Fields that no rule reads and that the evidence axis does not compare.} Examples are the
framework name, the SDK version, session and trace identifiers and the HTTP method.
\item \textbf{Parser hardening that the cases cannot exercise.} Removing the duplicate-key check or the
rejection of NaN and infinity from the strict JSON parser in the egress body projection survives,
because a case is a parsed dictionary and cannot contain a duplicate key or a non-finite number. This is
the territory of parser differentials, to which dictionary-level cases are blind by construction.
\item \textbf{A field carried twice.} The egress proxy sends the pinned address both as a resolved
address and as a range of one address. Deleting either survives, and the PDP resolves the host itself as
well.
\item \textbf{Two real gaps in what the packaged cases reach.} Deleting the authority source, and
deleting the delegation fields from the MCP gateway's request. \Cref{sec:r-fesource} describes them.
\end{enumerate}

\subsection{Engine mutants}

Of the \NmOpaSurv{} surviving Rego mutants, \NmOpaSurvRoleKnown{} delete a guard on a known role.
Where the body of the rule reads the role's configuration, Rego's undefined values make the guard
redundant, and the mutant is equivalent. Where it does not, namely for credential expiry, delegation
depth, a missing authority source and chain length, the mutant would report an extra failing rule for an
unknown role combined with that violation. No case contains that combination.

Most of the other surviving Rego mutants remove a test for a null or empty value that the following line
already implies, so they are equivalent. The remainder point at untested boundaries: the exact limits on
path length and segment count, an empty default list of blocked paths, and a chain whose length differs
from its depth. The cases that exercise the last one also need a reason that only the PDP produces, and
the engine-mutant runs excluded them.

The shared-helper survivors (\NmSharedSurv{} of \NmSharedTotal{}) are in the argument-path helpers (JSON
equality and the path walker), the address parser and the path normaliser. Both engines consume the
helper's output, and no expectation pins the behaviour.

\end{document}